\documentclass[twocolumn]{aastex63}

\usepackage{epsfig,natbib}
\usepackage{graphicx}
\usepackage{amsmath}
\usepackage{booktabs}
\usepackage{longtable}
\usepackage{xspace}
\usepackage[T1]{fontenc}
\usepackage{lipsum}
\newcommand{\K}{\ensuremath{\mathrm{K}}\xspace}

\def\deg{\ensuremath{^{\circ}}}

\shortauthors{MacDougall et al.}
\shorttitle{HIP-97166}
\begin{document}

\title{The TESS-Keck Survey. VI. Two Eccentric sub-Neptunes Orbiting HIP-97166}

\author[0000-0003-2562-9043]{Mason G.\ MacDougall}
\affiliation{Department of Physics \& Astronomy, University of California Los Angeles, Los Angeles, CA 90095, USA}

\author[0000-0003-0967-2893]{Erik A.\ Petigura}
\affiliation{Department of Physics \& Astronomy, University of California Los Angeles, Los Angeles, CA 90095, USA}

\author[0000-0002-9751-2664]{Isabel Angelo}
\affiliation{Department of Physics \& Astronomy, University of California Los Angeles, Los Angeles, CA 90095, USA}

\author[0000-0001-8342-7736]{Jack Lubin}
\affiliation{Department of Physics \& Astronomy, University of California Irvine, Irvine, CA 92697, USA}

%% TKS

\author[0000-0002-7030-9519]{Natalie M. Batalha}
\affiliation{Department of Astronomy and Astrophysics, University of California, Santa Cruz, CA 95060, USA}

\author[0000-0001-7708-2364]{Corey Beard}
\affiliation{Department of Physics \& Astronomy, University of California Irvine, Irvine, CA 92697, USA}

\author[0000-0003-0012-9093]{Aida Behmard}
\altaffiliation{NSF Graduate Research Fellow}
\affiliation{Division of Geological and Planetary Science, California Institute of Technology, Pasadena, CA 91125, USA}

\author{Sarah Blunt}
\altaffiliation{NSF Graduate Research Fellow}
\affiliation{Department of Astronomy, California Institute of Technology, Pasadena, CA 91125, USA}

\author{Casey Brinkman}
\altaffiliation{NSF Graduate Research Fellow}
\affiliation{Institute for Astronomy, University of Hawai`i, Honolulu, HI 96822, USA}

\author[0000-0003-1125-2564]{Ashley Chontos}
\altaffiliation{NSF Graduate Research Fellow}
\affiliation{Institute for Astronomy, University of Hawai`i, Honolulu, HI 96822, USA}

\author[0000-0002-1835-1891]{Ian J. M. Crossfield}
\affiliation{Department of Physics \& Astronomy, University of Kansas, Lawrence, KS 66045, USA}

\author[0000-0002-8958-0683]{Fei Dai}
\affiliation{Division of Geological and Planetary Science, California Institute of Technology, Pasadena, CA 91125, USA}

\author[0000-0002-4297-5506]{Paul A. Dalba}
\altaffiliation{NSF Astronomy and Astrophysics Postdoctoral Fellow}
\affiliation{Department of Earth and Planetary Sciences, University of California, Riverside, CA 92521, USA}

\author[0000-0001-8189-0233]{Courtney Dressing}
\affiliation{Department of Astronomy, University of California Berkeley, Berkeley, CA 94720, USA}

\author[0000-0003-3504-5316]{Benjamin Fulton}
\affiliation{NASA Exoplanet Science Institute/Caltech-IPAC, Pasadena, CA 91125, USA}

\author[0000-0002-8965-3969]{Steven Giacalone}
\affiliation{Department of Astronomy, University of California Berkeley, Berkeley, CA 94720, USA}

\author{Michelle L. Hill}
\affiliation{Department of Earth and Planetary Sciences, University of California, Riverside, CA 92521, USA}

\author[0000-0001-8638-0320]{Andrew W.\ Howard}
\affiliation{Department of Astronomy, California Institute of Technology, Pasadena, CA 91125, USA}

\author[0000-0001-8832-4488]{Daniel Huber}
\affiliation{Institute for Astronomy, University of Hawai`i, Honolulu, HI 96822, USA}

\author[0000-0002-0531-1073]{Howard Isaacson}
\affiliation{Department of Astronomy, University of California Berkeley, Berkeley, CA 94720, USA}
\affiliation{Centre for Astrophysics, University of Southern Queensland, Toowoomba, QLD, Australia}

\author[0000-0002-7084-0529]{Stephen R. Kane}
\affiliation{Department of Earth and Planetary Sciences, University of California, Riverside, CA 92521, USA}

\author{Andrew Mayo}
\affiliation{Department of Astronomy, University of California Berkeley, Berkeley, CA 94720, USA}

\author[0000-0003-4603-556X]{Teo Mo\v{c}nik}
\affiliation{Gemini Observatory/NSF's NOIRLab, 670 N. A'ohoku Place, Hilo, HI 96720, USA}

\author[0000-0001-8898-8284]{Joseph M. Akana Murphy}
\altaffiliation{NSF Graduate Research Fellow}
\affiliation{Department of Astronomy and Astrophysics, University of California, Santa Cruz, CA 95060, USA}

\author[0000-0001-7047-8681]{Alex Polanski}
\affiliation{Department of Physics \& Astronomy, University of Kansas, Lawrence, KS 66045, USA}

\author{Malena Rice}
\altaffiliation{NSF Graduate Research Fellow}
\affiliation{Department of Astronomy, Yale University, New Haven, CT 06520, USA}

\author[0000-0003-0149-9678]{Paul Robertson}
\affiliation{Department of Physics \& Astronomy, University of California Irvine, Irvine, CA 92697, USA}

\author[0000-0001-8391-5182]{Lee J.\ Rosenthal}
\affiliation{Department of Astronomy, California Institute of Technology, Pasadena, CA 91125, USA}

\author[0000-0001-8127-5775]{Arpita Roy}
\affiliation{Space Telescope Science Institute, Baltimore, MD 21218, USA}
\affiliation{Department of Physics and Astronomy, Johns Hopkins University, Baltimore, MD 21218, USA}

\author[0000-0003-3856-3143]{Ryan A. Rubenzahl}
\altaffiliation{NSF Graduate Research Fellow}
\affiliation{Department of Astronomy, California Institute of Technology, Pasadena, CA 91125, USA}

\author[0000-0003-3623-7280]{Nicholas Scarsdale}
\affiliation{Department of Astronomy and Astrophysics, University of California, Santa Cruz, CA 95060, USA}

\author{Emma Turtelboom}
\affiliation{Department of Astronomy, University of California Berkeley, Berkeley, CA 94720, USA}

\author{Judah Van Zandt}
\affiliation{Department of Physics \& Astronomy, University of California Los Angeles, Los Angeles, CA 90095, USA}

\author[0000-0002-3725-3058]{Lauren M. Weiss}
\affiliation{Institute for Astronomy, University of Hawai`i, Honolulu, HI 96822, USA}

% Gemini Data

\author{Elisabeth Matthews}
\affiliation{Observatoire de l'Universit\'{e} de Gen\`{e}ve, Chemin Pegasi 51, 1290 Versoix, Switzerland}

% TESS Architects

\author[0000-0002-4715-9460]{Jon M.\ Jenkins}
\affiliation{NASA Ames Research Center, Moffett Field, CA 94035-0001, USA}

\author[0000-0001-9911-7388]{David W.\ Latham}
\affil{Center for Astrophysics | Harvard \& Smithsonian, Cambridge, MA 02138, USA}

\author[0000-0003-2058-6662]{George R.\ Ricker}
\affil{Department of Physics and Kavli Institute for Astrophysics and Space Research, MIT, Cambridge, MA 02139, USA}

\author[0000-0002-6892-6948]{S. Seager}
\affil{Department of Earth, Atmospheric, and Planetary Sciences, MIT, Cambridge, MA 02139, USA}
\affil{Department of Physics and Kavli Institute for Astrophysics and Space Research, MIT, Cambridge, MA 02139, USA}
\affil{Department of Aeronautics and Astronautics, Massachusetts Institute of Technology, Cambridge, MA 02139, USA}

\author[0000-0001-6763-6562]{Roland K.\ Vanderspek}
\affiliation{Department of Physics and Kavli Institute for Astrophysics and Space Research, MIT, Cambridge, MA 02139, USA}

\author[0000-0002-4265-047X]{Joshua N.\ Winn}
\affiliation{Department of Astrophysical Sciences, Princeton University, Princeton, NJ 08544, USA}

% SPOC, POC, TSO nominees

\author[0000-0002-9314-960X]{C. E. Brasseur}
\affiliation{Space Telescope Science Institute, Baltimore, MD 21218, USA}

\author[0000-0002-9314-960X]{John Doty}
\affiliation{Noqsi Aerospace Ltd, 2822 S Nova Rd, Pine, CO 80470}

\author[0000-0002-9113-7162]{Michael Fausnaugh}
\affil{Department of Physics and Kavli Institute for Astrophysics and Space Research, MIT, Cambridge, MA 02139, USA}

\author{Natalia Guerrero}
\affil{Department of Physics and Kavli Institute for Astrophysics and Space Research, MIT, Cambridge, MA 02139, USA}

\author{Chris Henze}
\affiliation{NASA Ames Research Center, Moffett Field, CA 94035-0001, USA}

\author[0000-0003-2527-1598]{Michael B.\ Lund}
\affiliation{NASA Exoplanet Science Institute/Caltech-IPAC, Pasadena, CA 91125, USA}

\author[0000-0002-1836-3120]{Avi~Shporer}
\affiliation{Department of Physics and Kavli Institute for Astrophysics and Space Research, MIT, Cambridge, MA 02139, USA}

\begin{abstract}
We report the discovery of HIP-97166b (TOI-1255b), a transiting sub-Neptune on a 10.3-day orbit around a K0 dwarf 68 pc from Earth. This planet was identified in a systematic search of TESS Objects of Interest for planets with eccentric orbits, based on a mismatch between the observed transit duration and the expected duration for a circular orbit. We confirmed the planetary nature of HIP-97166b with ground-based radial velocity measurements and measured a mass of $M_{b} =$ 20 $\pm$ 2 $M_\earth$ along with a radius of $R_{b} =$ 2.7 $\pm$ 0.1 $R_\earth$ from photometry. We detected an additional non-transiting planetary companion with $M_{c}$ sin$i =$ 10 $\pm$ 2 $M_\earth$ on a 16.8-day orbit. While the short transit duration of the inner planet initially suggested a high eccentricity, a joint RV-photometry analysis revealed a high impact parameter $b = 0.84 \pm 0.03$ and a moderate eccentricity. Modeling the dynamics with the condition that the system remain stable over $>$10$^5$ orbits yielded eccentricity constraints $e_b = 0.16 \pm 0.03$ and $e_c < 0.25$. The eccentricity we find for planet b is above average for the small population of sub-Neptunes with well-measured eccentricities. We explored the plausible formation pathways of this system, proposing an early instability and merger event to explain the high density of the inner planet at $5.3 \pm 0.9$ g/cc as well as its moderate eccentricity and proximity to a 5:3 mean-motion resonance. 
\end{abstract}

\section{Introduction}
\label{sec:intro}

One of the key features of the Solar System is its low dynamical temperature. The eight planets are arranged with wide orbital spacing, low eccentricities, and no significant mean-motion resonances. Even so, the low mean eccentricity of $\sim$0.06 within the Solar System had to arise from somewhere. One explanation involves dynamical excitation and subsequent eccentricity damping following the divergent migration of Jupiter and Saturn (e.g. \citealt{Tsiganis05}). Given a different set of initial conditions, however, it is possible for this and other excitation processes to achieve even higher dynamical temperatures in other planetary systems.

One of the major surprises of early exoplanet observations was the prevalence of high-eccentricity Jovians, in direct contrast to the Solar System planets. For reference, $\sim$25$\%$ of known planets with $M_{p} > 100$ $M_\earth$ and $a > 1$ AU have $e > 0.4$ (NASA Exoplanet Archive; \citealt{Akeson13}). Various mechanisms have been proposed to explain the highly excited states of these Jovian orbits, including planet migration, resonances, and close approaches (see, \citealt{Fabrycky07}; \citealt{Ford08}; \citealt{Winn15}). 

On the other hand, characterizing the orbits and eccentricities of sub-Jovians is more challenging. The standard method of measuring a planet's eccentricity through radial velocity (RV) observations relies on the detection of a significant, non-sinusoidal motion from the host star. While many sub-Jovian semi-amplitudes are detectable with current facilities, the departures from sinusoidal are often less clear. Of the other available methods for planet detection, the transit method is the most prolific to date, but precise eccentricities are typically achieved through transit-timing variations (TTVs) when planets are near resonance, which is not representative of all systems (see, e.g., \citealt{Hadden14}). This limitation has led to a low fraction of sub-Neptune discoveries with well-constrained eccentricities.

\begin{figure}[ht]
\centering
\includegraphics[width=0.45\textwidth]{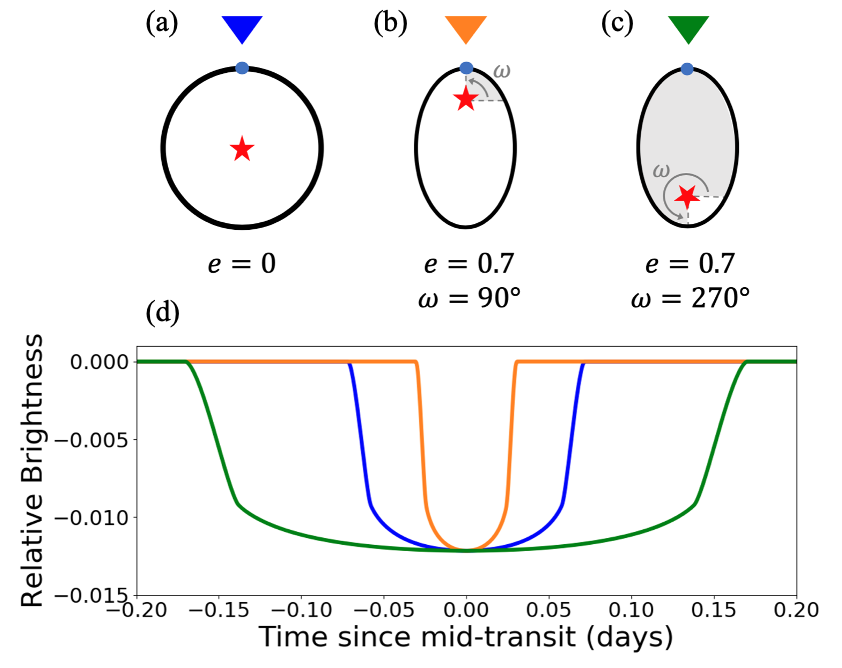}
\caption{Transit duration as a function of $e$ and $\omega$. (a) Observer (blue triangle) viewing a planet transiting on a circular orbit (blue light curve). (b) Observer (orange triangle) viewing a planet transiting at periastron (orange light curve). (c) Observer (green triangle) viewing a planet transiting at apastron (green light curve). (d) Simulated light curves associated with scenarios a-c for a Jupiter-size planet orbiting edge-on around a Sun-like star with a period of 5 days.}
\label{fig:e-w_dur}
\end{figure}

Fortunately, a connection between transit duration and eccentricity exists (see, e.g., \citealt{Ford08}), assuming one has a well-constrained estimate of impact parameter. This relationship has given rise to a variety of studies that derive dynamical insights from transit photometry alone (\citealt{Kipping10}; \citealt{Kipping14}; \citealt{Xie16}). Such work has been successfully carried out in recent years to determine both the eccentricities of individual planets (\citealt{Dawson12}; \citealt{VE14}) and the eccentricity distributions of larger samples of planetary systems (\citealt{Kane12}; \citealt{VE15}; \citealt{Xie16}). In our ongoing study, we use this relationship as a pre-filter to identify planet candidates at the extremes of the eccentricity distribution, observed with the Transiting Exoplanet Survey Satellite (TESS; \citealt{Ricker15}).

One of such candidates is HIP-97166b (TOI-1255b, TIC 237222864b), a sub-Neptune around an early K dwarf located high in the Northern hemisphere (dec = $+74\deg$) at a distance of 68 pc (Gaia Collaboration; \citealt{Lindegren18}). Our investigation is part of a larger effort by the TESS-Keck Survey (see \cite{Chontos21} for more information on TKS and its goals), which will build upon the legacies of Kepler and K2 to address major outstanding questions about exoplanet compositions, atmospheres, and system architectures.

\begin{figure*}[ht]
\centering
\includegraphics[width=0.95\textwidth]{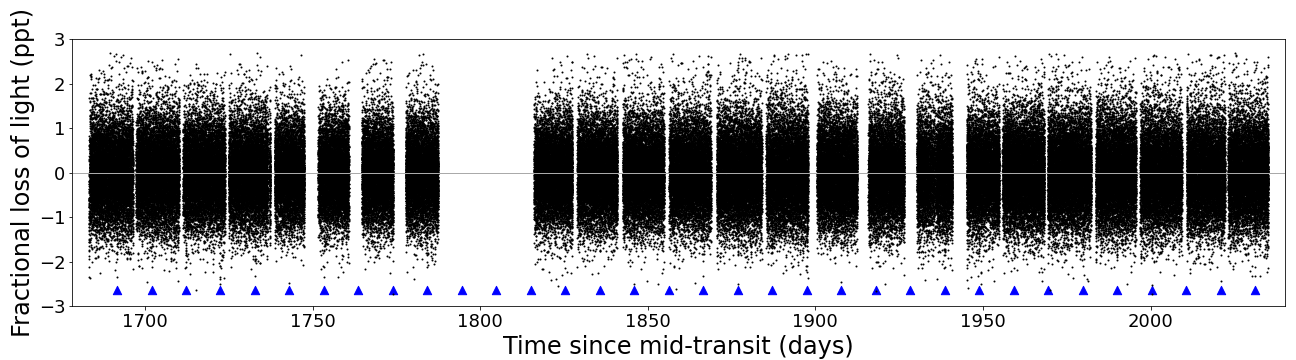}
\caption{TESS photometry of HIP-97166. The data has been processed following the procedures discussed in \S\ref{sec:photometry}. Transits of HIP-97166b are indicated by blue triangles and are 0.8 ppt.}
\label{fig:full_lc}
\end{figure*}

In this paper, we describe the HIP-97166 system and the transit profile modeling that we used to identify eccentric planet candidates from photometry (\S\ref{sec:ecc-candidate}) as well as our follow-up radial velocity observations (\S\ref{sec:rvs}). We analyze our spectroscopic measurements to characterize both the stellar (\S\ref{sec:stellar}) and planetary properties. From our rich RV data set, we also detect the presence of a non-transiting outer planet (\S\ref{sec:keplerian-model}). Finally, we explore the dynamics of this system through N-body simulations, which we used to further constrain our eccentricity measurements (\S\ref{sec:dynamics}). We also place this system in context with past exoplanet discoveries (\S\ref{sec:context}) and consider possible formation pathways that could have led to the observed system architecture (\S\ref{sec:formation}).

\section{HIP-97166\lowercase{b}: A High-Eccentricity Candidate}
\label{sec:ecc-candidate}

\subsection{TESS Photometry}
\label{sec:photometry}

HIP-97166 was observed by TESS with 2-min-cadence photometry in 12 sectors between UT 2019 July 18 and 2020 June 9 (Sectors 14--17, 19--26). The time-series photometry was processed by the TESS Science Processing Operations Center pipeline (SPOC; \citealt{Jenkins16}), which first detected the periodic transit signal of HIP-97166b in 2019 September with a wavelet-based, noise-compensating matched filter (\citealt{Jenkins02}; \citealt{Jenkins10}). An initial limb-darkened transit model fit was performed (\citealt{Li19}) and the signature passed a suite of diagnostic tests (\citealt{Twicken18}) but for the difference imaging centroid test, which located the source of the transit-like signal $10\farcs{}2 \pm 3\farcs{}1$ from the target. As the data accumulated for this target, the difference imaging centroid test results improved, shrinking the maximum deviation from the known target star's location to $0\farcs{}9 \pm 3\farcs{}5$ in the analysis of the data from sectors 14 through 26. The TESS Science Office reviewed the vetting results and issued an alert to the community on UT 2019 October 17 (\citealt{Guerrero21}). 

We accessed the Pre-search Data Conditioning Simple Aperture Photometry (PDC-SAP; \citealt{Stumpe12}, \citealt{Stumpe14}; \citealt{Smith12}) through the Mikulski Archive for Space Telescopes (MAST), stitching together the light curves from individual TESS sectors into a single time-series using \texttt{Lightkurve} (\citealt{lightkurve18}). We cleaned the TESS photometry by keeping only points with $quality\_flag=0$ and excluding outliers beyond 10-$\sigma$ from the baseline flux. We divided out the median background flux of the time-series data to normalize the light curve then searched for transits using a box least squares (BLS; \citealt{Kovacs02}) transit search to recover the same planetary signal detected by SPOC with SNR = 29.6. We subtracted the known transits and applied the BLS search again but identified no additional periodic transit events.

With the measured transit mid-point $t_0$ and period $P$ of the transiting planet, we masked out all transit events to detrend the light curve without obscuring the signal. We do note some photometric variability in the Simple Aperture Photometry (SAP) light curve, which we discuss in brief in \S\ref{sec:detection}. Interpolating over the masked transits, we fit a smoothed curve to systematics in the photometry with a Savitzky-Golay filter then subtracted out such additional structure to produce the flattened and normalized final light curve seen in Figure \ref{fig:full_lc}. Before unmasking the transit events, we clipped any individual outliers whose residuals to the smoothed fit were greater than 5-$\sigma$ discrepant.

\subsection{Photometric Transit Model}
\label{sec:transit-model}

The standard Mandel-Agol transit model can be specified by 5 transit parameters \{$P$, $t_0$, $R_p/R_*$, $b$, $a/R_*$\} in addition to stellar limb-darkening (\citealt{Mandel02}). Eccentricity $e$ and longitude of periastron $\omega$ can also be directly sampled by such models, but including these dynamical parameters can significantly increase model run-time. Before devoting extensive computation time to every TOI system, a faster calculation involving transit duration can serve as a pre-filter for planets with potentially eccentric orbits.

\begin{figure*}[ht]
\centering
\includegraphics[width=0.71\textwidth]{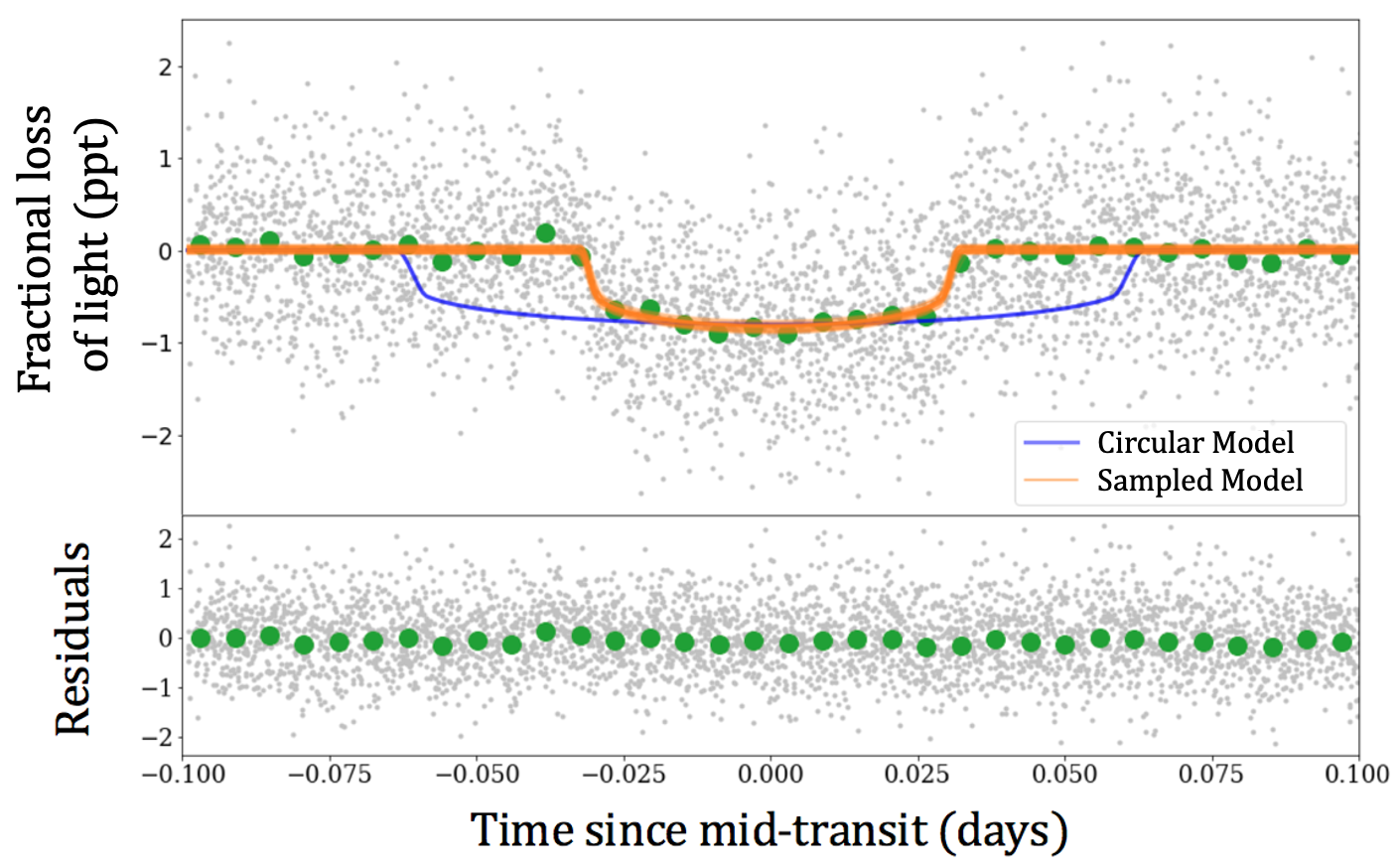}
\caption{Top: Transit models (orange; 50 samples) drawn from parameter posterior distributions fit from phase-folded TESS photometry of HIP-97166b. A simulated transit curve (blue) is shown for a theoretical circular orbit of HIP-97166b, modeled using median posterior distribution values for all other parameters. Details regarding fitting procedure are discussed in \S\ref{sec:transit-model}. Bottom: Residuals to our maximum \emph{a posteriori} model.}
\label{fig:transit-model}
\end{figure*}

While a planet on a circular orbit has a constant velocity, an eccentric planet will move faster when closer to its host star according to Kepler's 2nd Law. As a result, an observer viewing an eccentric planet transiting at periastron ($\omega_* = 90\deg$, see Figure \ref{fig:e-w_dur}b) would see a shorter duration while another observer viewing this transit at apastron ($\omega_* = 270\deg$, see Figure \ref{fig:e-w_dur}c) would see a longer duration. Thus, the ratio between a planet's observed transit duration and its duration if it were on a circular orbit depends on the eccentricity and orbital orientation. The true duration (mid-ingress to mid-egress) can be computed using the geometric relation given in \citealt{Winn10}:

\begin{equation}
\label{eqn:duration}
T = \left(\frac{R_* P}{\pi a}\sqrt{1-b^2}\right)\frac{\sqrt{1-e^2}}{1+e\sin{\omega}}.
\end{equation}

We modeled this effect by re-parametrizing the standard transit model to sample duration rather than $a/R_*$. Our fitting basis included \{$P$, $t_0$, $R_p/R_*$, $b$, $T$, $\mu$, $u$, $v$\}, where $\mu$ was mean out-of-transit stellar flux and $u$, $v$ were quadratic limb darkening parameters. We fit the transit photometry of HIP-97166 using the statistical transit modeling package \texttt{exoplanet} (\citealt{Foreman-Mackey2021}), which generates samples from the posterior probability density for these parameters conditioned on the observed TESS light curve. To generate these samples, \texttt{exoplanet} uses a gradient-based MCMC algorithm that is a generalization of the No U-Turn Sampling method (\citealt{Hoffman11}; \citealt{Betancourt16}). The versatility of \texttt{exoplanet} allowed us to build a model that best suited our goal here of obtaining a rapid transit duration fit. The sampled model produced a set of parameter posterior distributions from which we identified a median observed transit duration of $T_{obs} = 0.06$ days.

To calculate the theoretical "circular" transit duration $T_{circ}$ that the planet would have had given $e = 0$, we first point out that  $a/R_*$ in Eq. \ref{eqn:duration} maps to stellar density $\rho_*$ through Kepler's 3rd Law (assuming $m \ll M$):
%which we measured independently for HIP-97166 through the method discussed in \S\ref{sec:stellar}. 

%
\begin{equation}
\label{eqn:rho_substitution}
\frac{a}{R_*} = \left(\frac{P^2 G \rho_*}{3 \pi}\right)^{1/3}.
\end{equation}
Since $\rho_*$ can be directly measured through independent observations (see, e.g. \S\ref{sec:stellar}), a more useful parametrization of Eq. \ref{eqn:duration} is obtained by substituting in Eq. \ref{eqn:rho_substitution}, yielding
\begin{equation}
\label{eqn:duration_modified}
T \propto P^{1/3} \left(1-b^2\right)^{1/2} \rho_*^{-1/3} \frac{\sqrt{1-e^2}}{1+e\sin{\omega}}.
\end{equation}

Assuming one has an independently measured value for $\rho_*$, reliable estimates for $b$ and $P$, and $e = 0$, Eq. \ref{eqn:duration_modified} can be used to compute the expected transit duration of a planet if it were on a circular orbit (all else equal). For HIP-97166b, we found $T_{circ} = 0.12$ $\pm$ 0.02 days, for which a representative simulated transit curve can be seen in Figure \ref{fig:transit-model}. This calculation demonstrated that $T_{obs}$ is significantly shorter than $T_{circ}$, with $T_{obs}/T_{circ} \approx 0.5$, implying a potential highly eccentric orbit transiting near periastron.

\begin{figure}[ht]
\centering
\includegraphics[width=0.46\textwidth]{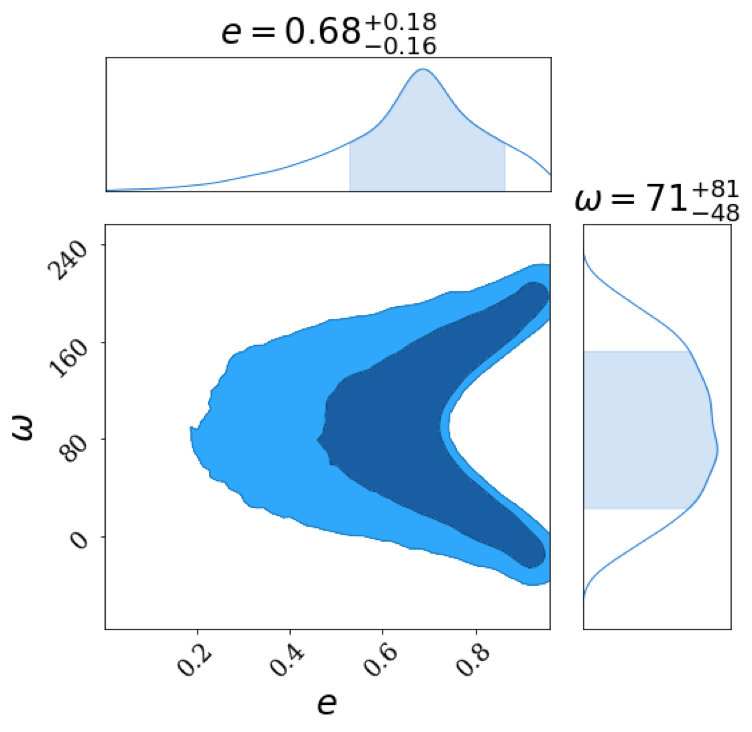}
\caption{2D joint posterior distribution of $e$ and $\omega$ for HIP-97166b, showing 1-$\sigma$ and 2-$\sigma$ credibility intervals. Best-fit values for both parameters are given: $e = 0.68\substack{+0.18 \\ -0.16}$ and $\omega = 71\deg\substack{+81\deg \\ -48\deg}$.}
\label{fig:e_w-posterior}
\end{figure}

Although each of the variables in our transit duration model represented a unique transit property, they were not completely independent. In particular, there is significant $e$-$\omega$-$b$ degeneracy for moderate-to-low SNR transits (\citealt{Petigura20}). For example, observed transit durations that are shorter than the expected duration for an edge-on, circular orbit can be caused by higher $b$ and/or higher $e$ (\citealt{Moorhead11}; \citealt{Dawson12}). On the other hand, this degeneracy is not a concern when modeling transits that are longer than expected since only eccentricity can have such an effect.

Given the short transit duration and these degeneracy concerns, we followed up our initial transit fit with a more detailed \texttt{exoplanet} model sampling 10 variables, \{$P$, $t_0$, $R_p/R_*$, $b$, $\rho_{*}$, $\sqrt{e}\sin{\omega}$, $\sqrt{e}\cos{\omega}$, $\mu$, $u$, $v$\}, each with weakly informative priors similar to those used by \cite{Sandford17}. We re-fit the TESS transit photometry of HIP-97166 with this model using 8,000 tuning steps and 6,000 sampling steps over 4 parallel chains, measuring an autocorrelation length of $\sim$8. We find that the largest Gelman-Rubin statistic amongst the sampled parameters is $R = 1.0004$, suggesting convergence of the posterior chains according to \cite{Gelman92}. Figure \ref{fig:transit-model} shows the final transit model sampled from the posteriors.

\subsection{Eccentricity Constraints from Photometry}
\label{sec:photo-ecc}

The photometrically-constrained eccentricity posterior distribution that we modeled for HIP-97166b was consistent with our initial high-eccentricity hypothesis, yielding a 1-$\sigma$ range of $e =$ 0.52--0.86. While $\omega$ is loosely constrained from this method, this analysis favors that the transit occurs closer to periastron. The joint 2D posterior for $e$ and $\omega$ is shown in Fig. \ref{fig:e_w-posterior}, with the 1-$\sigma$ and 2-$\sigma$ credible regions shown in shades of blue.

% Table of RV data
\begin{deluxetable}{lcrcc}
\tablewidth{0.99\textwidth}
\tabletypesize{\footnotesize}
\tablecaption{ Radial Velocity Measurements\label{tab:rv-data}}
\tablehead{
  \colhead{Time} & 
  \colhead{Tele.} &
  \colhead{RV} & 
  \colhead{RV Unc.} & 
  \colhead{$S_{HK}$} \\
  \colhead{(BJD)} & 
  \colhead{} & 
  \colhead{(m s$^{-1}$)} & 
  \colhead{(m s$^{-1}$)} & 
  \colhead{} 
}
\startdata
2458978.987&APF&0.549&3.407&0.202\\
2458979.905&APF&3.990&3.534&0.201\\
2458980.841&APF&-1.571&4.051&0.212\\
2458980.863&APF&-5.655&3.644&0.236\\
2458996.786&APF&5.818&3.092&0.278\\
2458996.807&APF&10.543&3.434&0.246\\
2459000.034&HIRES&-2.029&1.013&0.185\\
2459004.053&HIRES&-2.293&1.544&0.179\\
2459007.975&HIRES&-0.449&1.378&0.173\\
2459008.015&HIRES&-0.165&1.477&0.172\\
\enddata
\tablecomments{Only the first 10 RVs are displayed in this table. A complete list has been made available online. $S_{HK}$ values were measured using procedures from \cite{Isaacson10} with standard uncertainties of 0.002 for APF/Levy measurements and 0.001 for Keck/HIRES measurements.}
\end{deluxetable}

We note, however, that our impact parameter distribution remains loosely constrained as well, with a 1-$\sigma$ range of $b =$ 0.28--0.79, peaking in density towards the upper end of this range. Due to this high variance in the posterior distributions of both $b$ and $e$, we required additional observations to break the $e$-$\omega$-$b$ degeneracy of our transit model.

\section{Spectroscopic Follow-up}
\label{sec:rvs}

\subsection{HIRES RVs}
\label{sec:hires}

We collected 44 spectra of HIP-97166 with the HIRES instrument at the Keck Observatory (\citealt{Vogt94}) between UT 2020 May 30 and UT 2021 April 9 (Table \ref{tab:rv-data}). On average, the observations have a spectral resolution of $R$ = 50,000, using a median exposure time of 900 s at 5500 \AA{}. We also obtained a high-SNR template spectrum on UT 2020 June 27 with 400 SNR pixel$^{-1}$ at 5000 \AA{}.

\begin{deluxetable}{lrc}
\tablewidth{0.99\textwidth}
\tabletypesize{\footnotesize}
\tablecaption{HIP-97166 System Properties\label{tab:system-properties}}
\tablehead
{
  \multicolumn{1}{l}{Parameter}&
  \multicolumn{1}{r}{Value}&
  \multicolumn{1}{c}{Notes}
}

\startdata
$Stellar$   &   &   \\
RA (\deg)	& 296.24462	& A \\
Dec (\deg)	& 74.06286	& A \\
$\pi$ (mas)	& $15.134 \pm 0.023$	& A \\
$m_K$	& $7.92 \pm 0.02$	& B \\
$m_V$	& $9.92 \pm 0.03$	& C \\
$T_{eff}$ (K)	& $5198 \pm 100$	& D \\
$[$Fe/H$]$ (dex)	& $0.27 \pm 0.09$	& D \\
$log(g)$	& $4.41 \pm 0.10$	& D \\
$age$ (Gyr)	& $3.33 \pm 3.28$	 & E \\
$M_*$ ($M_\odot$)	& $0.898 \pm 0.054$	 & E \\
$R_*$ ($R_\odot$)	&	 $0.836 \pm 0.036$	& E \\
$\rho_*$ (g/cc)	& $2.154 \pm 0.312$	& E \\
$S_{HK}$	& $0.182$	& F \\
log$R'_{HK}$	& $-5.01$	& F \\
$u, v$	& $0.47 \pm 0.05$, $0.10 \pm 0.05$	 & G \\
\\
$Planet$ $b$    &   &   \\
$P$ (days)	& $10.28891 \pm 0.00004$	& H \\
$T_c$ (BJD-2457000)	& $1691.6486 \pm 0.0007$	& H \\
$b$	& $0.836 \pm 0.027$	 & H \\
$R_p$ ($R_\earth$)	& $2.74 \pm 0.13$	& H,E \\
$M_p$ ($M_\earth$)	& $20.0 \pm 1.5$	& I \\
$\rho_p$ (g/cc)	& $5.3 \pm 0.9$	& I \\
$a$ (AU)	&	 $0.089 \pm 0.001$	& I \\
$\omega$ ($\deg$)	& $120.9 \pm 15.5$	& I \\
$e$	& $0.16 \pm 0.03$	& J \\
$T_{eq}$ (K)	& $757 \pm 25$	& K \\
\\
$Planet$ $c$    &   &   \\
$P$ (days)	& $16.84 \pm 0.22$	& I \\
$T_c$ (BJD-2457000)	& $1988.4 \pm 1.6$	& I \\
$M_p$ sin$i$ ($M_\earth$)	& $9.9 \pm 1.8$	& I \\
$a$ (AU)	&	 $0.124 \pm 0.002$	& I \\
$\omega$ ($\deg$)	& $178.2 \pm 53.0$	& I \\
$e$	& $< 0.25$	& J \\
$T_{eq}$ (K)	& $642 \pm 22$	& K \\
\enddata
\tablecomments{A: \emph{Gaia} DR2 (\citealt{GaiaDR2}); B: \emph{2MASS} (\citealt{Skrutskie06}); C: TESS Input Catalog (TIC; \citealt{Stassun19}); D: Derived with \texttt{SpecMatch-Synth}; E: Derived with \texttt{isoclassify}; F: Measured from Keck/HIRES template; G: Derived with \texttt{LDTK} (\citealt{ldtk15}); H: Constrained from \texttt{exoplanet} transit model; I: Best-fit RV model with \texttt{RadVel}; J: Dynamically constrained with \texttt{rebound}; K: Derived.}
\end{deluxetable}

For such observations, a heated cell of gaseous iodine was included along the light path just behind the entrance slit of the spectrometer, imprinting a dense forest of molecular absorption lines onto the observed stellar spectrum (\citealt{Marcy92}). These lines served as a wavelength reference for measuring the relative Doppler shift of each spectrum and tracking variations in the instrument profile using the standard forward-modeling procedures of the California Planet Search (\citealt{Howard10}). Along with the measured RVs and corresponding uncertainties, the stellar activity S-index was computed for all 43 Keck/HIRES observations using the observed strengths of the Ca II H and K lines in our template spectrum, following the methods of \cite{Isaacson10}.

\subsection{APF RVs}
\label{sec:apf}

In addition to the Keck/HIRES follow-up, we also collected 124 iodine-in spectra of HIP-97166 with the Levy spectrograph on the Automated Planet Finder (APF) telescope (\citealt{Vogt14}) between UT 2020 May 9 and UT 2021 April 11 at a spectral resolution of $R$ = 100,000. For a majority of these, two observations were taken roughly 30 minutes apart over 70 separate nights. We also obtained 7 iodine-free template spectra using APF/Levy on UT 2020 May 26 with an average SNR pixel$^{-1}$ of 102 at 5100 \AA{}. 

The APF/Levy Doppler code was developed based on the Keck/HIRES Doppler code and therefore follows a similar process for reducing spectra to RVs. However, this target is close to the APF/Levy magnitude limit with $V = 9.85$, contributing to a high cosmic ray rate that ultimately rendered the APF/Levy template unusable. Fortunately, we were able to substitute this with the the Keck/HIRES template to extract APF RVs, which has been successfully done in previous studies (see, e.g. \citealt{Dai20}).

\section{Stellar Characterization}
\label{sec:stellar}

We searched for nearby stars to rule out any contamination scenarios which would dilute the transit depth and, therefore, underestimate the planet size. We note that the star has a single neighbor listed in \emph{Gaia} Data Release 2 (DR2) within 30\arcsec{} which, at $\Delta$G = 8.75, contributes negligible dilution (\citealt{GaiaDR2}). Although \emph{Gaia} is capable of spatially resolving sources down to $\sim$0\farcs{}5 separation in some instances, this still leaves a small region around our target star in which additional contaminating sources could exist. 

\begin{figure}[ht]
\centering
\includegraphics[width=0.45\textwidth]{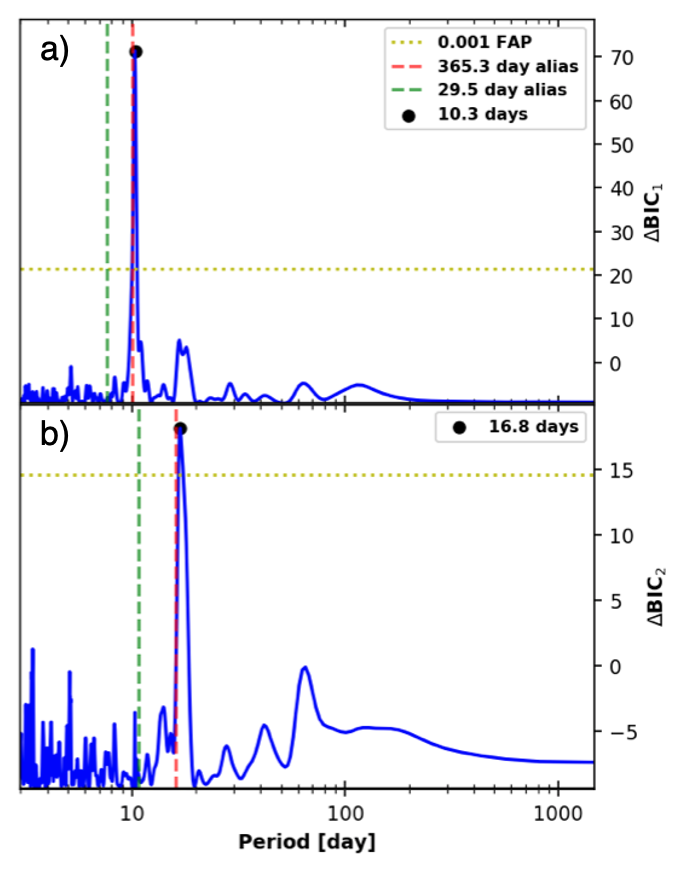}
\caption{Iterative Keplerian periodogram search of HIP-97166 RVs using \texttt{RVSearch}, confirming the 10.3-day transiting planet and identifying a significant non-transiting companion with a 16.8-day period. $\Delta$BIC is used to discriminate between models with Keplerians at varying periods (Bayesian Information Criterion; \citealt{Schwarz78}), corresponding to a significance threshold of FAP = 0.001 at the yellow dashed line. Monthly and yearly aliases are shown with in green and red dashed lines, respectively.}
\label{fig:rv-search}
\end{figure}

We used Gemini/NIRI adaptive optics imager (\citealt{Hodapp03}) to collect high resolution images of HIP-97166 on UT 2019 December 4. We collected nine images in the Br$\gamma$ filter, each with exposure time 1.75 s, in a grid dither pattern. We also collected flat frames and used the median-combined, dithered science frames as a sky background frame. For each frame we removed bad pixels, flat-fielded, and subtracted the sky background. We then aligned each image to the position of the frame and co-added the stack of images. We searched for companions visually, and did not detect companions anywhere in the field of view (26\farcs{}7$\times$26\farcs{}7, centered on the target). To assess the sensitivity of these observations, we injected fake PSFs at a number of position angles and separations from the host star and scaled the brightness of these such that they could be detected at 5$\sigma$. We reached a contrast of 5-mag relative to the host star beyond 270 mas and of 7.3-mag in the background limited regime beyond $\sim$1\farcs{}1. We were thus able to rule out close-in diluting sources with high certainty.

We sought to further characterize HIP-97166 by inferring $T_{eff}$ and [Fe/H] from our Keck/HIRES template spectrum using \texttt{SpecMatch-Synth}, as described in \cite{Petigura17b}. Following the methodology of \cite{FultonPetigura18}, we then used these values as priors for stellar isochrone modeling with \texttt{isoclassify} (\citealt{Berger20a}; \citealt{Huber17}). Our model also incorporated \emph{2MASS} K-band magnitude and \emph{Gaia} parallax to identify the best-fit stellar properties according to the MESA Isochrones and Stellar Tracks models (MIST; \citealt{Dotter16}; \citealt{Choi16a}). We characterized $\rho_*$ and several other stellar parameters using this method, accounting for model grid uncertainties according to \cite{Tayar20}, and we present these values in table \ref{tab:system-properties}.

\begin{figure}[ht]
\centering
\includegraphics[width=0.45\textwidth]{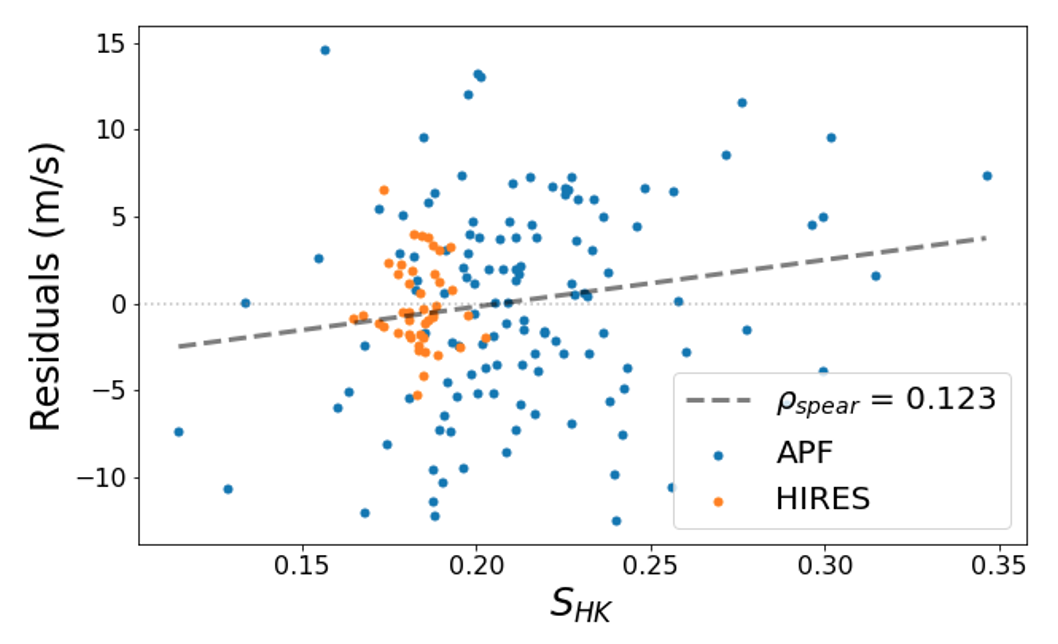}
\caption{Spearman rank-order linear correlation fit between 1-planet RV residuals and $S_{HK}$ activity metric shows a tenuous trend with a low correlation score $\rho_{spear}$, suggesting that no significant relationship exists between the observed RV signals and stellar activity.}
\label{fig:activity}
\end{figure}

Based on the solar-like values measured for $log(g)$, $S_{HK}$, and log$R'_{HK}$, HIP-97166 places among the bottom quartile of expected activity-induced RV jitter, according to \cite{Luhn20}. The minimum jitter of similar stars in this study is $\lesssim 2.5$ m/s, which is consistent with a jitter measurement of $\sim$2.5 m/s from our RVs. 

\section{Keplerian Modeling}
\label{sec:keplerian-model}

\subsection{RV Detection of Planets b and c }
\label{sec:detection}

A preliminary search of our RV data for periodic signals using \texttt{RVSearch} (\citealt{Rosenthal21}) revealed a Keplerian signal that matched the period of the transiting planet, with a false-alarm probability FAP $\ll$ 0.001 and a Doppler semi-amplitude of $K \approx 6.6$ m/s. After subtracting, we identified an additional signal at 16.8 days, with $K \approx 2.8$ m/s and FAP $\approx$ $10^{-3}$ (Figure \ref{fig:rv-search}).

We also searched for planets using the $l_1$ periodogram described in \citealt{Hara17}, designed to reduce periodogram noise for unevenly sampled data as compared to the Lomb-Scargle method by solving the Basis Pursuit minimization problem (\citealt{Chen01}). In our implementation, we used jitter term $\sigma = 2.5$ m/s, correlation time $\tau = 0$, and maximum frequency of 1.5 cycles d$^{-1}$. Within the period range of 1.1 to 1000 days, the only clear detections in the $l_1$-periodogram occurred near the known period of 10.3 days and within 1-$\sigma$ of the suspected period of 16.8 days, with FAP values of $\sim$$10^{-9}$ and $\sim$$10^{-3}$, respectively. 

To confirm the significance of the 16.8 day period relative to other plausible signals that could be achieved by random fluctuations in noise, we re-sampled $10^3$ synthetic data sets from the original RV data (\citealt{Howard10}). We found that the 16.8 day signal was consistently the next strongest signal found in $l_1$-periodogram searches of the synthetic data, with the 10.3 day period always being the most significant.

\begin{figure}[ht]
\centering
\includegraphics[width=0.45\textwidth]{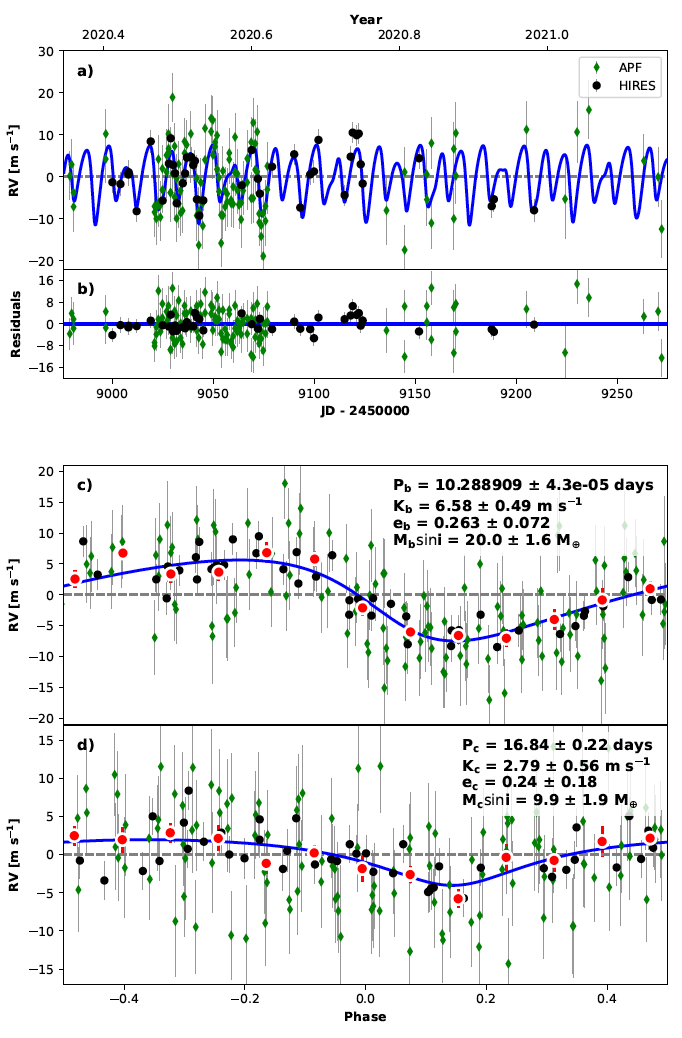}
\caption{a) Best-fit model (blue) of radial velocity measurements from Keck/HIRES (black) and APF/Levy (green) using \texttt{Radvel} (\citealt{Fulton18a}) with no binning; b) Residuals to best-fit RV model; c) Phase-folded view of best-fit model and RV data for HIP-97166b, with binned points shown in red; d) Phase-folded view of best-fit model and RV data for HIP-97166c}
\label{fig:radvel-model}
\end{figure}

While the 16.8-day RV signal was statistically significant, we did not identify a corresponding transit in \S\ref{sec:photometry}. We confirmed this by phase-folding the detrended TESS photometry according to the RV-constrained period $P$ and time of conjunction $T_c$ for the outer RV signal, detecting no evidence for a transit event at this period. We therefore considered the possibility that this signal was stellar activity induced. We searched for trends in the $S_{HK}$ activity time-series described in \S\ref{sec:stellar} and shown in Table \ref{tab:rv-data}. Similar to the RV data, we processed the nightly $S_{HK}$ measurements using an $l_1$-periodogram, showing no indication of stellar activity with a 16.8-day period nor any other statistically significant periodicity (FAP $\leq$ $10^{-3}$). Applying a similar approach to the SAP photometry of HIP-97166, we identified a significant $l_1$-periodogram signal at a period of 27.2 days, which was inconsistent with the 16.8 day RV signal. This periodicity in the SAP photometry is likely a systematic effect associated with the orbital period of the TESS spacecraft which has been seen in other TESS light curves as well.

To further test if the Keplerian signal was driven by stellar variability, we looked for a correlation between the $S_{HK}$ index time-series and the RV residuals after removing the 10.3-day planet signal. Using the Spearman rank-order correlation test (\citealt{Press92}), however, we found only a tenuous correlation with coefficient $\rho_{spear}$ = 0.12 (Figure \ref{fig:activity}). We, therefore, conclude that the 16.8-day periodicity is planetary in origin.

\subsection{RV-only Constraints}
\label{sec:rv-fit}

\begin{figure*}[ht]
\centering
\includegraphics[width=0.95
\textwidth]{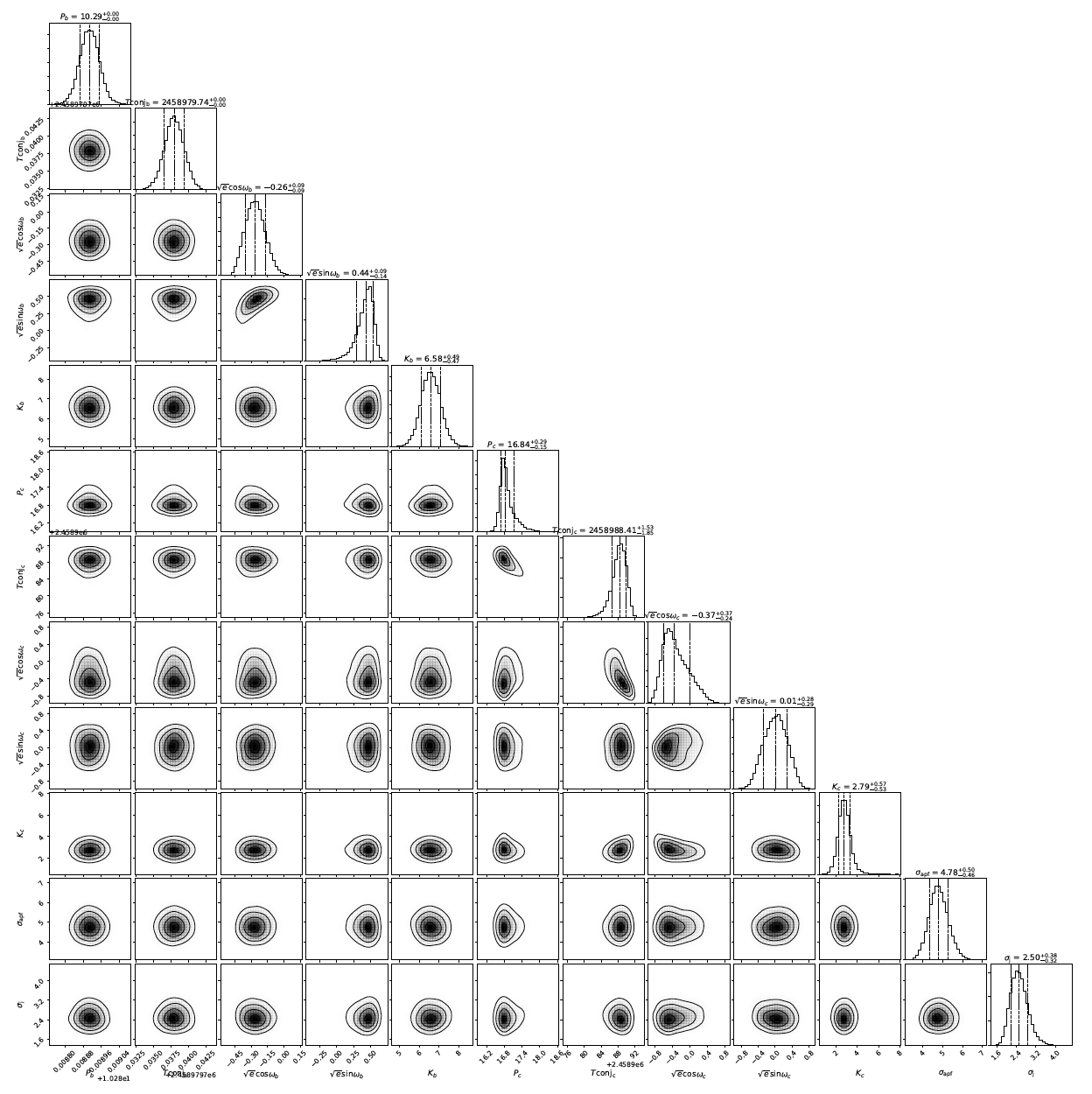}
\caption{Posterior distributions of parameters from two-planet RV-only model with \texttt{RadVel}.}
\label{fig:radvel-corner}
\end{figure*}

We fit our complete RV data set using a two-planet model with \texttt{RadVel}, a Python package that applies maximum \emph{a posteriori} model fitting and parameter estimation via MCMC to characterize planets from Keplerian RV signals (\citealt{Fulton18a}). The model that we selected consisted of the following free parameters for both planets: $P$, $T_c$, $K$, $\sqrt{e} \cos{\omega}$, and $\sqrt{e} \sin{\omega}$. We also included RV offsets $\gamma$ and RV jitter terms $\sigma$ for each instrument, accounting for other astrophysical and instrumental uncertainty that are not already included in the model. Figure \ref{fig:radvel-model} shows the best-fit two-planet model as determined by the posterior distributions for each parameter shown in Figure \ref{fig:radvel-corner}. 

Our best-fit model confirms the existence of two eccentric sub-Jovians orbiting HIP-97166, with a summary of planet properties provided in Table \ref{tab:system-properties}. Notably, the mass constraints for both planets b and c are significant, at the $\sim$10-$\sigma$ and $\sim$5-$\sigma$ levels, respectively. We also constrained the eccentricity of the transiting planet, $e_b = 0.26$ $\pm$ 0.07, which was $\sim$2.5-$\sigma$ below the median $e_b$ value of our transit model posterior distribution.

\subsection{RV-Photometry Joint Model}
\label{sec:joint-fit}

While the RV-measured eccentricity was lower than expected, our RV and photometric posterior distributions on $e_b$ remained consistent at the 2-$\sigma$ level. The degeneracy between $e$ and $b$ in our transit model was likely the source of this high spread, so we sought to build a complete model that more accurately accounted for this degeneracy. We performed a joint RV-transit fit of the data, using the posteriors on $e$ and $\omega$ from our RV fit as priors in a transit model to determine the combination of $e$ and $b$ needed to account for both the anomalously short transit duration and the RV signal.

In addition to $e$ and $\omega$ priors, we also placed a Gaussian prior on $\rho_*$ in our RV-informed transit model, informed by our \texttt{isoclassify} stellar characterization. All other aspects of the model remained the same from the earlier implementation of \texttt{exoplanet} in \S\ref{sec:transit-model}. These RV-derived priors allowed us to constrain impact parameter to $b = 0.84$ $\pm$ 0.03, significantly more constrained than our initial model fit and more consistent with what we observed in our RV-derived eccentricity posterior. Due to the covariance between $b$ and $R_p/R_*$, a higher $b$ also meant a higher $R_p/R_*$ and subsequently resulted in our final planet radius measurement of $R_p = 2.74$ $\pm$ 0.13 $R_\earth$.

\section{System Dynamics}
\label{sec:dynamics}

\subsection{Eccentricity Constraints from Stability Requirements}
\label{sec:stability}

The best-fit RV model suggested a significant (>3-$\sigma$) non-zero eccentricity for the inner transiting planet but had only limited constraints for the eccentricity of the outer planet. Given this dynamical assessment and the relative proximity of these orbits ($\Delta$a $\approx$ 0.035 AU $\approx$ 30 $\Delta_{Hill}$), orbit crossing constraints revealed that the best-fit $e_b$ and $e_c$ values existed in an unstable region of parameter space. We subsequently evaluated the long-term stability and effects of planet-planet interactions within this system. We applied dynamical constraints on the orbital properties of both planets using N-body code \texttt{rebound} (\citealt{Rein12}), initializing 10,000 orbital simulations of the HIP-97166 system with properties drawn from our joint model posterior distributions.

For each simulation, we randomly drew the various system parameters ($M_*$, $M_b$, $M_c$, $P_b$, $P_c$, $\omega_b$, $\omega_c$) from the posterior distributions of our joint model. For $e_b$ and $e_c$, we performed a similar random sampling but with conditions that prevented orbit crossing at the initial state of the system based on derived values of $a_b$ and $a_c$. We ran the simulations for up to $10^4$ years ($\sim$$3 \times $$10^5$ orbits of planet b) and found that only $\sim$33\% of credible models were stable.

There is a region of $e_b$-$e_c$ parameter space where orbital stability is preferred that is consistent within $\sim$1.5-$\sigma$ of the values derived from the best-fit RV model. Simulations that successfully completed the $10^4$ year run had overall eccentricity distributions given by $e_b = 0.16$ $\pm$ 0.03 and $e_c < 0.25$. Although the best-fit $e$ values from our RV model are within the upper tail of the dynamically constrained distributions, it is clear that our \texttt{RadVel} model posteriors, which do not account for orbit crossing, skew towards higher values than allowed by stability criteria. Nonetheless, we confirmed that HIP-97166b can maintain a moderate eccentricity over long timescales even with a mildly eccentric, nearby companion.

\begin{figure}[ht]
\centering
\includegraphics[width=0.45\textwidth]{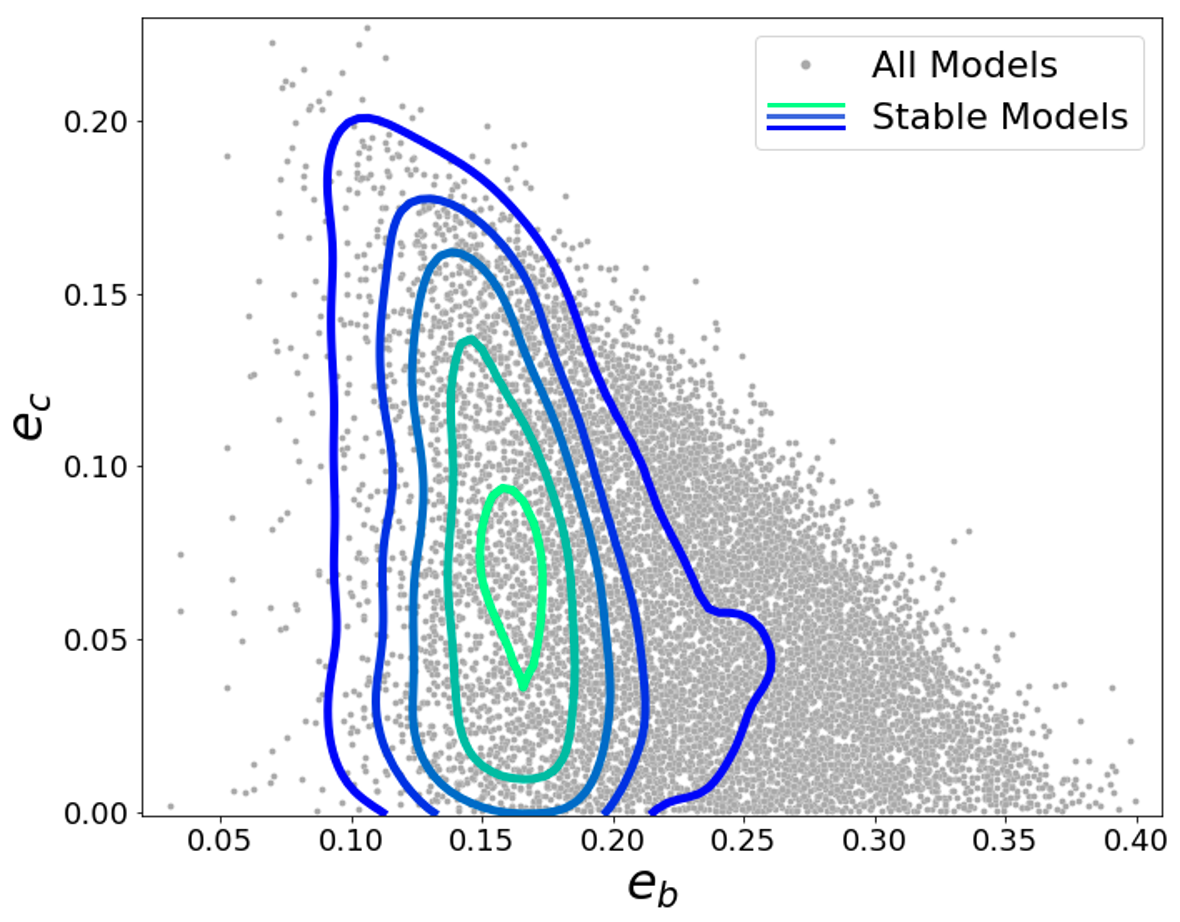}
\caption{Distribution of initialized $e_b$-$e_c$ values for all \texttt{rebound} simulations (grey) of the HIP-97166 system. Blue-green contours show regions with the highest density of stable eccentricity configurations (simulations lasting $10^4$ years).}
\label{fig:rebound}
\end{figure}

\subsection{Secular Eccentricity Variability}
\label{sec:secular}

\begin{figure*}[ht]
\centering
\includegraphics[width=0.95\textwidth]{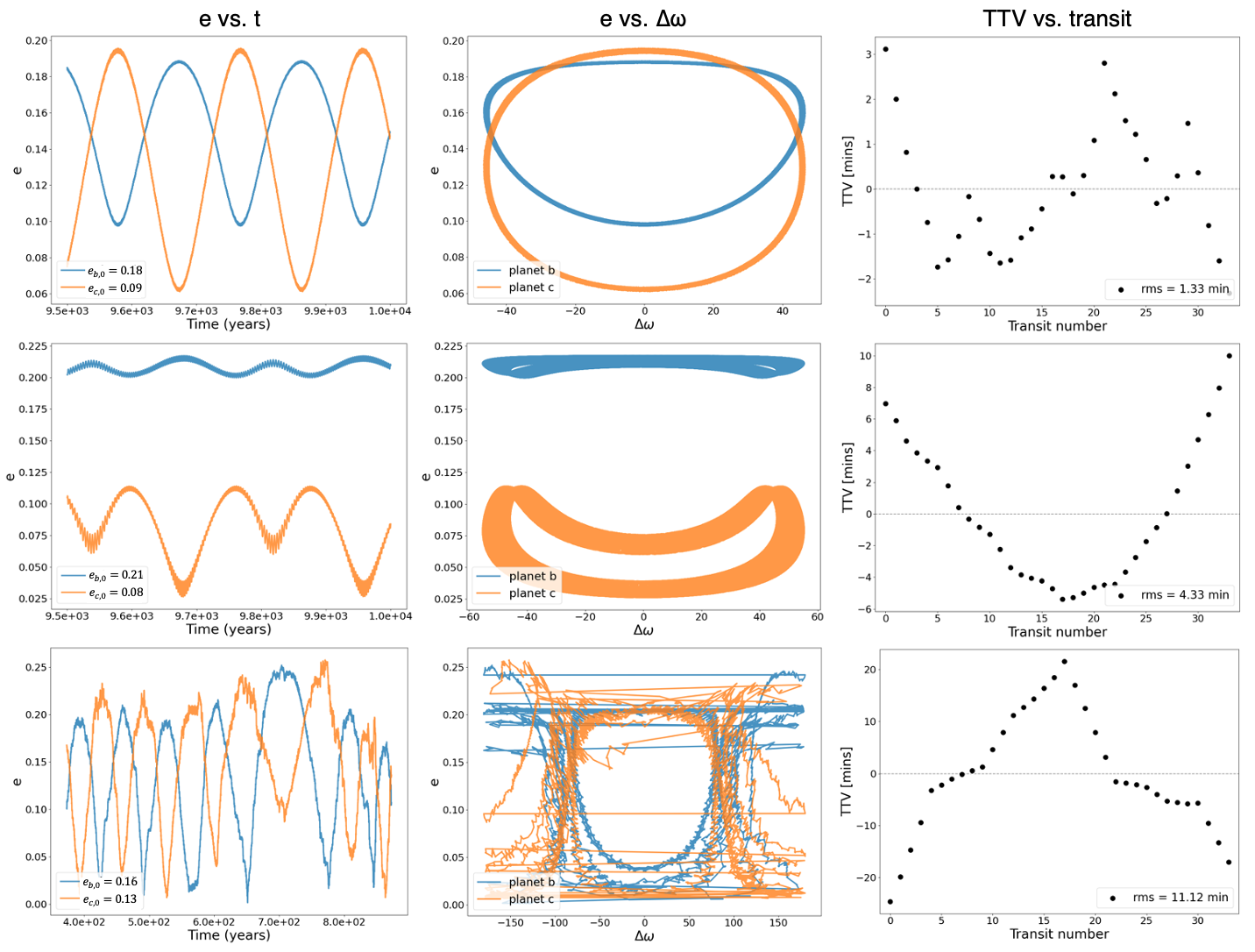}
\caption{Three different dynamical scenarios of HIP-97166 b and c eccentricities and TTVs simulated using \texttt{rebound} and sampled from our previous results as described in \S\ref{sec:secular}. Rows: Stable eccentric scenario with negligible TTVs (Top). Stable eccentric scenario with periods oscillating about 5:3 MMR, displaying non-negligible TTVs (Middle). Unstable eccentric scenario lasting only $\sim$800 years with moderate TTVs (Bottom). Columns: $e$ as a function of time over final 500 years of integration (Left). $e$ as a function of $\Delta\omega$ libration (Middle). TTV amplitude as a function of transit number over the same time baseline as TESS photometry (Right).}
\label{fig:secular}
\end{figure*}

For simulations that experienced orbit crossing prior to $10^4$ years, significant exchanges between $e_b$ and $e_c$ drove the system towards the instability region of $e_b$-$e_c$ parameter space. Stable systems, on the other hand, exhibited Laplace-Lagrange oscillations that remained stable long-term with a secular timescale on the order of $\sim$10$^2$ years (Figure \ref{fig:secular}). While neither planet's eccentricity is expected to reach exceptionally high values during such oscillations, it is interesting to consider which formation scenarios could have led to this compact, excited orbital architecture. We note that these orbits are near 5:3 second-order mean-motion resonance (MMR), $P_c/P_b$ $\approx$ 1.64 $\pm$ 0.02, but only in rare cases did we observe an impact of this resonance on the long-term stability of our simulations (see Figure \ref{fig:secular}, middle row). Although we cannot confirm the existence of MMR in this system, we cannot rule this scenario out either given the uncertainty on $P_b$.

\subsection{TTV Analysis}
\label{sec:ttvs}

Through our dynamical simulations, we also computed the expected magnitude of the TTVs experienced by the inner planet due to interactions from the outer planet in the HIP-97166 system (Figure \ref{fig:secular}, 3rd column). Evaluating the RMS of these simulated TTVs over the same baseline as our TESS photometry, we found that stable simulations had a TTV RMS distribution of 1.8 $\pm$ 1.8 mins. We also measured the magnitude of any observed TTVs from the transit photometry using \texttt{exoplanet}, finding an O-C RMS of $\sim$2.9 mins. This agreement between the simulated and empirical TTV RMS values demonstrated a consistency between the photometry and our proposed dynamics in which neither detected a significant TTV signal.

\section{System in Context}
\label{sec:context}

\subsection{Bulk Density and Core-Envelope Fraction}
\label{sec:composition}

HIP-97166b is a sub-Neptune, a class of planets with radii $\sim$2--4 $R_\earth$ that have been the subject of numerous studies in recent years (\citealt{Marcy14}; \citealt{Weiss14}; \citealt{Lopez14a}; \citealt{Wolfgang16}). Sub-Neptunes display a wide range of characteristics and compositions but tend to have densities lower than that of Earth, suggesting H/He envelopes of a few percent (see, e.g., \citealt{Rogers14}; \citealt{Owen17}). With respect to the \cite{Weiss14} relation, HIP-97166b is an outlier in the mass-radius parameter space, with bulk properties that place it among the denser and more massive planets in this class like K2-110b (\citealt{Osborn17}).

\begin{figure*}[ht]
\centering
\includegraphics[width=0.95\textwidth]{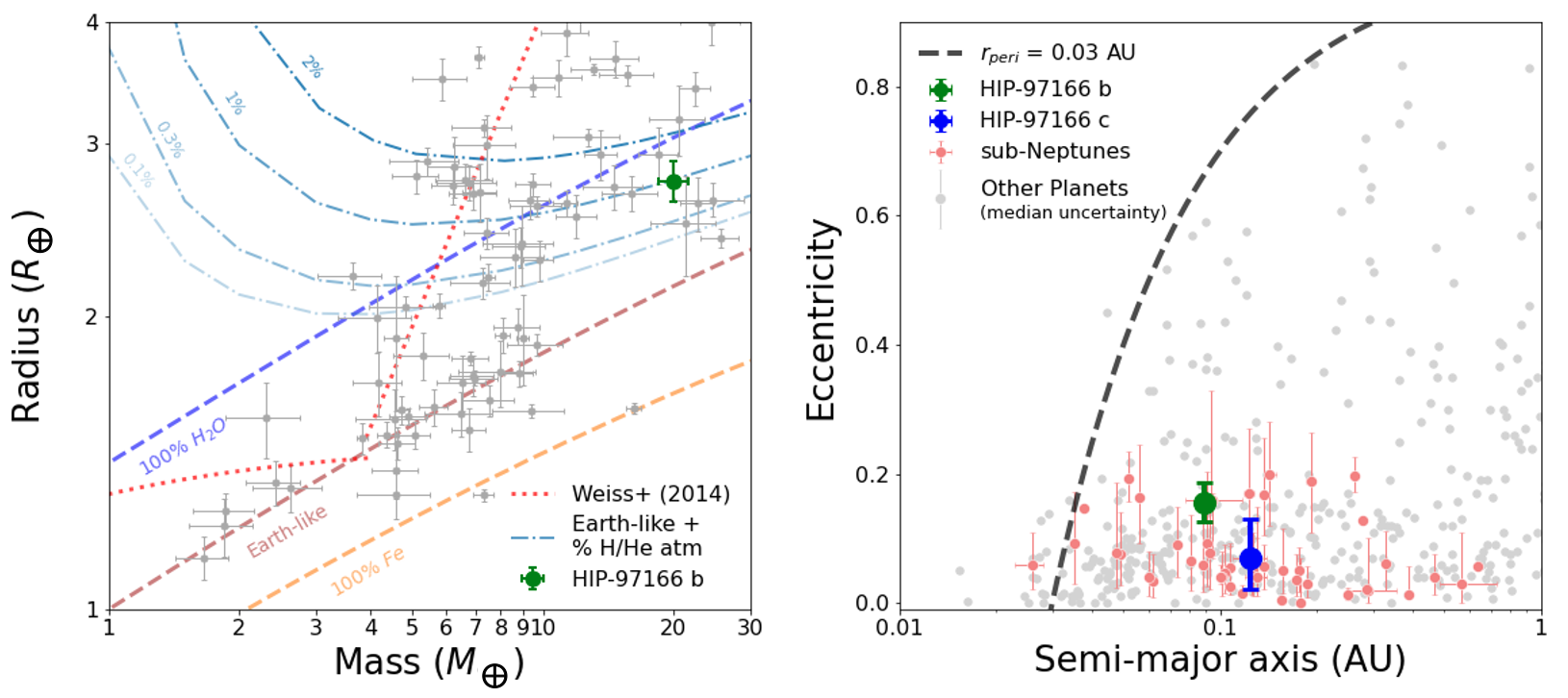}
\caption{Left: Mass-radius distribution of known sub-Neptunes (\citealt{Akeson13}) with 20$\%$ measurement precision or better in both mass and radius (grey), shown with HIP-97166b (green). Planetary interior composition curves for various 2-component models (no atmosphere) from \cite{Zeng16} are shown as dashed colored lines. Composition curves which include $H_2$ envelopes of varying size on top of an Earth-like core are given by faded blue dot-dash lines (\citealt{Zeng19}), assuming an equilibrium temperature of 700 K. \cite{Weiss14} model shown as red dotted line (note the log-log axes). Right: Eccentricity distribution of planets with $\sigma_e$ < 0.1 (sub-Neptunes in red, other known planets in gray) as a function of orbital separation, showing HIP-97166 b and c in green and blue, respectively. While the eccentricity constraints of both planets place their upper bounds towards the eccentric tail of the sub-Neptune distribution, they are consistent with overall suppressed dynamical temperatures of sub-Neptunes relative to other planet populations. Periastron distance of 0.03 AU is also shown.}
\label{fig:context}
\end{figure*}

We investigated the composition of HIP-97166b using a 2-component model, following the procedures of \cite{Petigura17} to quantify the core and envelope fractions of this planet. Assuming an Earth-like core composition and a solar-composition H/He envelope, we interpolated over a 4D grid of planetary and stellar parameters to quantitatively derive an estimate of the envelope mass fraction using \cite{Lopez14a} planet structure models. We identified an envelope fraction of 1.4 $\pm$ $0.4\%$, giving a core mass of $\sim$19.6 $M_\earth$ (see Figure \ref{fig:context}).

As for the outer planet, it is difficult to comment on its composition without a detectable transit. In the event that the outer planet is indeed transiting but below the detection threshold (SNR < 8), we could place an upper limit on the radius of planet c. Assuming a transit duration of 0.1 days and no inclination, we find that $R_{p,c} \leq$ 1.5 $R_{\earth}$. A high-density planet of this size with expected mass $\sim$10 $M_{\earth}$ would be an outlier in $M_{p}-R_{p}$ space, but it cannot necessarily be ruled out. A sufficiently inclined orbit, however, could allow nearly any planet size, so we are left with only a minimum mass measurement at this time.

\subsection{Weak Eccentricity Damping}
\label{sec:damping}

While we found no indication of eccentricity decay in this system during our 10,000 year simulations, we considered the extent of tidal circularization on longer timescales following the procedure of \cite{Petigura17}. \cite{Goldreich66} give the timescale for tidal eccentricity damping:

\begin{equation}
\label{eqn:tides}
\tau_e = \frac{4}{63}\left(\frac{Q'}{n}\right)\left(\frac{M_P}{M_*}\right)\left(\frac{a}{R_P}\right)^5
\end{equation}

The mean motion and the reduced tidal quality factor are given by $n = \sqrt{G M_* / a^3}$ and $Q' = 3 Q / 2 k_2$, respectively, where $Q$ is the specific dissipation function and $k_2$ is the tidal Love number (\citealt{Goldreich66}; \citealt{Murray99}; \citealt{Mardling04}). While $Q'$ is highly uncertain even for Solar System planets, we estimated its value for the sub-Neptune HIP-97166b based on the range of values associated with Earth ($\sim10^3$ -- $2\times10^3$) and Uranus ($\sim10^5$ -- $6\times10^5$), drawn from \cite{Lainey16} and \cite{Petigura17}. Assuming an Earth-like $Q' = 10^3$ from the lower end of this range, we calculated $\tau_e \approx 5$ Gyr. This was a similar timescale to the expected age of the system ($3.3\substack{+4.1 \\ -2.3}$ Gyr) based on our \texttt{isoclassify} model, implying that significant eccentricity decay was unlikely to have occurred so far in this scenario. Since $\tau_e$ scales linearly with $Q'$, larger values of $Q'$ would only continue to inflate $\tau_e$. Thus, we conclude that the observed eccentricity of HIP-97166b is, at most, susceptible only to weak tidal damping over large timescales. 

\subsection{Sub-Neptune Eccentricity}
\label{sec:eccentricity}

To date, sub-Neptunes make up 29$\%$ of all planet discoveries, but only 6.4$\%$ of planets with eccentricity constraints of $\sigma_e <$ 0.1 fall within this radius range (\citealt{Akeson13}). In Figure \ref{fig:context}, we show HIP-97166 b and c in context with other sub-Neptunes with well-constrained eccentricities. The small sample size makes it difficult to characterize the underlying eccentricity distribution of sub-Neptunes. Nevertheless, current observations suggest suppressed dynamical temperatures among this population, including the system presented in this paper. While our RV fit showed a 1-$\sigma$ eccentricity range of $e_b =$ 0.19 -- 0.33, our dynamically constrained N-body model lowered this 1-$\sigma$ range to $e_b =$ 0.13 -- 0.19. We display the latter result in Figure \ref{fig:context}, adopting this as our final characterization. This finding is consistent with that of \cite{Correia20}, which suggests that warm Neptune-mass planets tend to present moderate, non-zero eccentricities.

\cite{VE15} also carried out an investigation into the eccentricity distribution of small planets using a sample of \emph{Kepler} multi-planet systems. The eccentricity measurements in this study were performed using similar methods as we employed in our work in \S\ref{sec:ecc-candidate}. The authors found that the overall eccentricity distribution was consistent with a Rayleigh function with dispersion $\sigma_e =$ 0.049 $\pm$ 0.013, which indicated that smaller planets ($R_{p} \sim$ 2.6 $R_{\earth}$) generally had lower eccentricities when in the presence of planetary companions. A follow-up study investigating both multi-planet and single-planet transiting systems implemented a different distribution parametrization but still found that small planets in multi systems had lower eccentricities (\citealt{VE19}). The modeled distributions for multi and single planet systems in this work were consistent with half-Gaussians with dispersions $\sigma_e =$ 0.083 $\pm$ 0.018 and $\sigma_e =$ 0.32 $\pm$ 0.06.

Given these past findings, we assert that HIP-97166b has a typical eccentricity relative to other single transiting systems of small planets. Similarly, the eccentricity of HIP-97166b is also consistent within $\sim$1-$\sigma$ of the typical well-characterized sub-Neptune. We do note, however, that in comparison to well-characterized Jovian planets, sub-Neptunes display a trend of suppressed eccentricities verified by a Kolmogorov-Smirnov test. The discovery of HIP-97166b is consistent with the observed trend.

\section{Formation Scenarios}
\label{sec:formation}

Although past studies have shown that \emph{Kepler} multi-planet systems display no preference for being near mean-motion resonances (\citealt{Fabrycky14}), resonances may still play a role in shaping some system architectures. One possible formation pathway for the HIP-97166 system is convergent migration, a process of interactions with smaller bodies in the circumstellar disk that lowers the period ratio and often drives planets into resonant configurations (\citealt{Ford08}). While it has been noted that 5:3 MMR is a poor configuration for long-term stability (\citealt{Lee09}), it is possible that the presently observed 5:3 near-MMR is a result of the past crossing of a stronger resonance. The dynamical instability and subsequent scattering from this compact resonance would be sufficient to excite the eccentricities that we observed for these two planets (\citealt{Chiang02}; \citealt{Izidoro17}). 

While the HIP-97166 system is $<2\%$ away from the 5:3 second-order resonance, it is also only $\sim18\%$ away from the stronger 2:1 resonance, making this a plausible source of a past dynamical instability as well. A complication with this hypothesis, however, is that convergent migration requires dissipation and is expected to naturally damp eccentricities over time. While it is possible that the free eccentricities of HIP-97166 b and c may have been larger in the past if the orbits were spaced further apart, we also consider other possible explanations.

Another formation pathway for the HIP-97166 system is that the eccentricities were excited as the result of dynamical instabilities from planet-planet scattering and/or merger events (see, e.g., \citealt{Juric08}). \cite{Chiang12} demonstrated that in-situ formation of close-in super-Earths and sub-Neptunes can result in planet-planet mergers that lead to more massive planets on compact orbits. This led us to suggest that both the high density and moderate eccentricity of HIP-97166b may be a consequence of a past merger event between two $\sim$10 $M_\earth$ planets. The occurrence of three or more planets of similar size on evenly spaced, compact orbits like this is not unheard of and has previously been referred to as the "peas in a pod" effect (\citealt{Weiss18}). An instability in the dynamics of either inner planet in this proposed origin scenario could have easily resulted in a merged $\sim$20 $M_\earth$ planet with $e \sim$ 0.2, exciting the orbit of the outer planet in the process.

Finally, we considered the possibility that the observed planetary orbits may have been excited by a distant giant planetary companion. A giant perturber would have the capacity to increase inner planet eccentricities through dynamical interactions (\citealt{Hansen17}; \citealt{Becker17}; \citealt{Pu18}) or complex resonance effects such as the Eccentric Kozai-Lidov Mechanism (\citealt{Naoz16}; \citealt{Denham19}; \citealt{Barnes20}). Our RV observations presented only a marginal detection of acceleration, which allowed us to place loose constraints on a possible distant giant companion. With an observational baseline of $\sim$12-months, we computed that an unobserved Jupiter-mass companion or larger could have a separation as low as a few AU but larger separations were more likely. These results indicated that the distant giant excitation scenario could not entirely be ruled out.

\section{Summary and Conclusions}
\label{sec:conclusions}

In this work, we identified HIP-97166b as an eccentric sub-Neptune candidate. This is the first in a series of investigations into TOIs with transit durations that significantly differ from expectations of a circular orbit. Combining our duration pre-filter and transit model with an analysis of follow-up RV observations from Keck/HIRES and APF/Levy, we measured the mass, radius, and moderate eccentricity of HIP-97166b. This sub-Neptune is both denser ($\rho_{p,b} = 5.3 \pm 0.9$ g/cc) and more eccentric ($e_b = 0.16 \pm 0.03$) than is typical for planets of similar size, making it an interesting find among the TESS candidates. We also discovered a moderately eccentric outer companion ($e_c < 0.25$) from RV observations, with a minimum mass of 10 $M_\earth$ and a 16.8 day orbit. 

N-body simulations of these orbits over time revealed a narrow region of dynamical stability that allowed us to measure the eccentricity of the inner planet with high precision, excluding a circular orbit to $\sim$5-$\sigma$. Our leading hypothesis is that this system originally formed with a "peas in a pod" architecture, where the inner two of three original planets merged after a dynamical instability placed them on crossing orbits. The eccentricities that we observed in this system could have resulted from such an event and persisted on timescales of the age of the system. HIP-97166b is now among a small group of sub-Neptunes with high-precision eccentricity measurements.

\begin{acknowledgments}
The authors thank the anonymous reviewer and data
editor for constructive comments on the manuscript. MM acknowledges support from the UCLA Cota-Robles Graduate Fellowship. DH acknowledges support from the Alfred P. Sloan Foundation, the National Aeronautics and Space Administration (80NSSC21K0652), and the National Science Foundation (AST-1717000). PD is supported by a National Science Foundation (NSF) Astronomy and Astrophysics Postdoctoral Fellowship under award AST-1903811. The authors also thank Daniel Foreman-Mackey for discussions regarding the parametrization of the transit models used in this work.

This work was supported by a NASA Keck PI Data Award, administered by the NASA Exoplanet Science Institute. Data presented herein were obtained at the W. M. Keck Observatory from telescope time allocated to the National Aeronautics and Space Administration through the agency's scientific partnership with the California Institute of Technology and the University of California. The Observatory was made possible by the generous financial support of the W. M. Keck Foundation.

We thank the time assignment committees of the University of California, the California Institute of Technology, NASA, and the University of Hawaii for supporting the TESS-Keck Survey with observing time at Keck Observatory and on the Automated Planet Finder.  We thank NASA for funding associated with our Key Strategic Mission Support project.  We gratefully acknowledge the efforts and dedication of the Keck Observatory staff for support of HIRES and remote observing.  We recognize and acknowledge the cultural role and reverence that the summit of Maunakea has within the indigenous Hawaiian community. We are deeply grateful to have the opportunity to conduct observations from this mountain.  

We thank Ken and Gloria Levy, who supported the construction of the Levy Spectrometer on the Automated Planet Finder. We thank the University of California and Google for supporting Lick Observatory and the UCO staff for their dedicated work scheduling and operating the telescopes of Lick Observatory.  This paper is based on data collected by the TESS mission. Funding for the TESS mission is provided by the NASA Explorer Program. This paper also includes data that are publicly available from the Mikulski Archive for Space Telescopes (MAST).

We acknowledge the use of public TESS data from pipelines at the TESS Science Office and at the TESS Science Processing Operations Center. Resources supporting this work were provided by the NASA High-End Computing (HEC) Program through the NASA Advanced Supercomputing (NAS) Division at Ames Research Center for the production of the SPOC data products. This work also used computational and storage services associated with the Hoffman2 Shared Cluster provided by UCLA Institute for Digital Research and Education's Research Technology Group.

Part of this work is based on observations obtained at the international Gemini Observatory, a program of NSF's NOIRLab, which is managed by the Association of Universities for Research in Astronomy (AURA) under a cooperative agreement with the National Science Foundation. On behalf of the Gemini Observatory partnership: the National Science Foundation (United States), National Research Council (Canada), Agencia Nacional de Investigaci\'{o}n y Desarrollo (Chile), Ministerio de Ciencia, Tecnolog\'{i}a e Innovaci\'{o}n (Argentina), Minist\'{e}rio da Ci\^{e}ncia, Tecnologia, Inova\c{c}\~{o}es e Comunica\c{c}\~{o}es (Brazil), and Korea Astronomy and Space Science Institute (Republic of Korea). Data were collected under program GN-2019B-LP-101.
\end{acknowledgments}

\facilities{TESS, Keck/HIRES, APF/Levy, \emph{Gaia}, \emph{Gemini}/NIRI}

\software{We made use of the following publicly available Python modules: \texttt{exoplanet} (\citealt{Foreman-Mackey2021}), \texttt{PyMC3} (\citealt{pymc16}), \texttt{theano} (\citealt{theano16}), \texttt{LDTK} (\citealt{ldtk15}), \texttt{RadVel} (\citealt{Fulton18a}), \texttt{RVSearch} (\citealt{Rosenthal21}), \texttt{astropy} (\citealt{astropy:2013}, \citealt{astropy:2018}), \texttt{isoclassify} (\citealt{Huber17}), \texttt{lightkurve} (\citealt{lightkurve18}), \texttt{chainconsumer} (\citealt{Hinton16}), \texttt{matplotlib} (\citealt{Hunter07}), \texttt{numpy} (\citealt{harris2020array}), \texttt{scipy} (\citealt{SciPy20}), and \texttt{pandas} (\citealt{pandas}).}

\bibliographystyle{aasjournal}
\bibliography{adslib}

\begin{thebibliography}{}
\expandafter\ifx\csname natexlab\endcsname\relax\def\natexlab#1{#1}\fi
\providecommand{\url}[1]{\href{#1}{#1}}

\bibitem[{{Akeson} {et~al.}(2013){Akeson}, {Chen}, {Ciardi}, {Crane}, {Good},
  {Harbut}, {Jackson}, {Kane}, {Laity}, {Leifer}, {Lynn}, {McElroy}, {Papin},
  {Plavchan}, {Ram{\'\i}rez}, {Rey}, {von Braun}, {Wittman}, {Abajian}, {Ali},
  {Beichman}, {Beekley}, {Berriman}, {Berukoff}, {Bryden}, {Chan}, {Groom},
  {Lau}, {Payne}, {Regelson}, {Saucedo}, {Schmitz}, {Stauffer}, {Wyatt}, \&
  {Zhang}}]{Akeson13}
{Akeson}, R.~L., {Chen}, X., {Ciardi}, D., {et~al.} 2013, \pasp, 125, 989

\bibitem[{{Astropy Collaboration} {et~al.}(2013){Astropy Collaboration},
  {Robitaille}, {Tollerud}, {Greenfield}, {Droettboom}, {Bray}, {Aldcroft},
  {Davis}, {Ginsburg}, {Price-Whelan}, {Kerzendorf}, {Conley}, {Crighton},
  {Barbary}, {Muna}, {Ferguson}, {Grollier}, {Parikh}, {Nair}, {Unther},
  {Deil}, {Woillez}, {Conseil}, {Kramer}, {Turner}, {Singer}, {Fox}, {Weaver},
  {Zabalza}, {Edwards}, {Azalee Bostroem}, {Burke}, {Casey}, {Crawford},
  {Dencheva}, {Ely}, {Jenness}, {Labrie}, {Lim}, {Pierfederici}, {Pontzen},
  {Ptak}, {Refsdal}, {Servillat}, \& {Streicher}}]{astropy:2013}
{Astropy Collaboration}, {Robitaille}, T.~P., {Tollerud}, E.~J., {et~al.} 2013,
  \aap, 558, A33

\bibitem[{{Astropy Collaboration} {et~al.}(2018){Astropy Collaboration},
  {Price-Whelan}, {Sip{\H{o}}cz}, {G{\"u}nther}, {Lim}, {Crawford}, {Conseil},
  {Shupe}, {Craig}, {Dencheva}, {Ginsburg}, {Vand erPlas}, {Bradley},
  {P{\'e}rez-Su{\'a}rez}, {de Val-Borro}, {Aldcroft}, {Cruz}, {Robitaille},
  {Tollerud}, {Ardelean}, {Babej}, {Bach}, {Bachetti}, {Bakanov}, {Bamford},
  {Barentsen}, {Barmby}, {Baumbach}, {Berry}, {Biscani}, {Boquien}, {Bostroem},
  {Bouma}, {Brammer}, {Bray}, {Breytenbach}, {Buddelmeijer}, {Burke},
  {Calderone}, {Cano Rodr{\'\i}guez}, {Cara}, {Cardoso}, {Cheedella}, {Copin},
  {Corrales}, {Crichton}, {D'Avella}, {Deil}, {Depagne}, {Dietrich}, {Donath},
  {Droettboom}, {Earl}, {Erben}, {Fabbro}, {Ferreira}, {Finethy}, {Fox},
  {Garrison}, {Gibbons}, {Goldstein}, {Gommers}, {Greco}, {Greenfield},
  {Groener}, {Grollier}, {Hagen}, {Hirst}, {Homeier}, {Horton}, {Hosseinzadeh},
  {Hu}, {Hunkeler}, {Ivezi{\'c}}, {Jain}, {Jenness}, {Kanarek}, {Kendrew},
  {Kern}, {Kerzendorf}, {Khvalko}, {King}, {Kirkby}, {Kulkarni}, {Kumar},
  {Lee}, {Lenz}, {Littlefair}, {Ma}, {Macleod}, {Mastropietro}, {McCully},
  {Montagnac}, {Morris}, {Mueller}, {Mumford}, {Muna}, {Murphy}, {Nelson},
  {Nguyen}, {Ninan}, {N{\"o}the}, {Ogaz}, {Oh}, {Parejko}, {Parley}, {Pascual},
  {Patil}, {Patil}, {Plunkett}, {Prochaska}, {Rastogi}, {Reddy Janga},
  {Sabater}, {Sakurikar}, {Seifert}, {Sherbert}, {Sherwood-Taylor}, {Shih},
  {Sick}, {Silbiger}, {Singanamalla}, {Singer}, {Sladen}, {Sooley},
  {Sornarajah}, {Streicher}, {Teuben}, {Thomas}, {Tremblay}, {Turner},
  {Terr{\'o}n}, {van Kerkwijk}, {de la Vega}, {Watkins}, {Weaver}, {Whitmore},
  {Woillez}, {Zabalza}, \& {Astropy Contributors}}]{astropy:2018}
{Astropy Collaboration}, {Price-Whelan}, A.~M., {Sip{\H{o}}cz}, B.~M., {et~al.}
  2018, \aj, 156, 123

\bibitem[{{Barnes} {et~al.}(2020){Barnes}, {Haswell}, {Staab},
  {Anglada-Escud{\'e}}, {Fossati}, {Doherty}, {Cooper}, {Jenkins}, {D{\'\i}az},
  {Soto}, \& {Pe{\~n}a Rojas}}]{Barnes20}
{Barnes}, J.~R., {Haswell}, C.~A., {Staab}, D., {et~al.} 2020, Nature
  Astronomy, 4, 419

\bibitem[{{Becker} \& {Adams}(2017)}]{Becker17}
{Becker}, J.~C., \& {Adams}, F.~C. 2017, \mnras, 468, 549

\bibitem[{{Berger} {et~al.}(2020){Berger}, {Huber}, {van Saders}, {Gaidos},
  {Tayar}, \& {Kraus}}]{Berger20a}
{Berger}, T.~A., {Huber}, D., {van Saders}, J.~L., {et~al.} 2020, \aj, 159, 280

\bibitem[{{Betancourt}(2016)}]{Betancourt16}
{Betancourt}, M. 2016, arXiv e-prints, arXiv:1601.00225

\bibitem[{{Chen} {et~al.}(2001){Chen}, {Donoho}, \& {Saunders}}]{Chen01}
{Chen}, S.~S., {Donoho}, D.~L., \& {Saunders}, M.~A. 2001, SIAM Review, 43, 129

\bibitem[{{Chiang} \& {Laughlin}(2013)}]{Chiang12}
{Chiang}, E., \& {Laughlin}, G. 2013, \mnras, 431, 3444

\bibitem[{{Chiang} {et~al.}(2002){Chiang}, {Fischer}, \& {Thommes}}]{Chiang02}
{Chiang}, E.~I., {Fischer}, D., \& {Thommes}, E. 2002, \apjl, 564, L105

\bibitem[{{Choi} {et~al.}(2016){Choi}, {Dotter}, {Conroy}, {Cantiello},
  {Paxton}, \& {Johnson}}]{Choi16a}
{Choi}, J., {Dotter}, A., {Conroy}, C., {et~al.} 2016, \apj, 823, 102

\bibitem[{{Chontos} {et~al.}(2021){Chontos}, {Akana Murphy}, {MacDougall},
  {Fetherolf}, {Van Zandt}, {Rubenzahl}, {Beard}, {Huber}, {Batalha},
  {Crossfield}, {Dressing}, {Fulton}, {Howard}, {Isaacson}, {Kane}, {Petigura},
  {Robertson}, {Roy}, {Weiss}, {Behmard}, {Dai}, {Dalba}, {Giacalone}, {Hill},
  {Lubin}, {Mayo}, {Mocnik}, {Polanski}, {Rosenthal}, {Scarsdale}, \&
  {Turtelboom}}]{Chontos21}
{Chontos}, A., {Akana Murphy}, J.~M., {MacDougall}, M.~G., {et~al.} 2021, arXiv
  e-prints, arXiv:2106.06156

\bibitem[{{Correia} {et~al.}(2020){Correia}, {Bourrier}, \&
  {Delisle}}]{Correia20}
{Correia}, A.~C.~M., {Bourrier}, V., \& {Delisle}, J.~B. 2020, \aap, 635, A37

\bibitem[{{Dai} {et~al.}(2020){Dai}, {Roy}, {Fulton}, {Robertson}, {Hirsch},
  {Isaacson}, {Albrecht}, {Mann}, {Kristiansen}, {Batalha}, {Beard}, {Behmard},
  {Chontos}, {Crossfield}, {Dalba}, {Dressing}, {Giacalone}, {Hill}, {Howard},
  {Huber}, {Kane}, {Kosiarek}, {Lubin}, {Mayo}, {Mocnik}, {Akana Murphy},
  {Petigura}, {Rosenthal}, {Rubenzahl}, {Scarsdale}, {Weiss}, {Van Zandt},
  {Ricker}, {Vanderspek}, {Latham}, {Seager}, {Winn}, {Jenkins}, {Caldwell},
  {Charbonneau}, {Daylan}, {G{\"u}nther}, {Morgan}, {Quinn}, {Rose}, \&
  {Smith}}]{Dai20}
{Dai}, F., {Roy}, A., {Fulton}, B., {et~al.} 2020, \aj, 160, 193

\bibitem[{{Dawson} \& {Johnson}(2012)}]{Dawson12}
{Dawson}, R.~I., \& {Johnson}, J.~A. 2012, \apj, 756, 122

\bibitem[{{Denham} {et~al.}(2019){Denham}, {Naoz}, {Hoang}, {Stephan}, \&
  {Farr}}]{Denham19}
{Denham}, P., {Naoz}, S., {Hoang}, B.-M., {Stephan}, A.~P., \& {Farr}, W.~M.
  2019, \mnras, 482, 4146

\bibitem[{{Dotter}(2016)}]{Dotter16}
{Dotter}, A. 2016, \apjs, 222, 8

\bibitem[{{Fabrycky} \& {Tremaine}(2007)}]{Fabrycky07}
{Fabrycky}, D., \& {Tremaine}, S. 2007, \apj, 669, 1298

\bibitem[{{Fabrycky} {et~al.}(2014){Fabrycky}, {Lissauer}, {Ragozzine}, {Rowe},
  {Steffen}, {Agol}, {Barclay}, {Batalha}, {Borucki}, {Ciardi}, {Ford},
  {Gautier}, {Geary}, {Holman}, {Jenkins}, {Li}, {Morehead}, {Morris},
  {Shporer}, {Smith}, {Still}, \& {Van Cleve}}]{Fabrycky14}
{Fabrycky}, D.~C., {Lissauer}, J.~J., {Ragozzine}, D., {et~al.} 2014, \apj,
  790, 146

\bibitem[{{Ford} \& {Rasio}(2008)}]{Ford08}
{Ford}, E.~B., \& {Rasio}, F.~A. 2008, \apj, 686, 621

\bibitem[{Foreman-Mackey {et~al.}(2021)Foreman-Mackey, Luger, Agol, Barclay,
  Bouma, Brandt, Czekala, David, Dong, Gilbert, Gordon, Hedges, Hey, Morris,
  Price-Whelan, \& Savel}]{Foreman-Mackey2021}
Foreman-Mackey, D., Luger, R., Agol, E., {et~al.} 2021, Journal of Open Source
  Software, 6, 3285.
\newblock \url{https://doi.org/10.21105/joss.03285}

\bibitem[{{Fulton} \& {Petigura}(2018)}]{FultonPetigura18}
{Fulton}, B.~J., \& {Petigura}, E.~A. 2018, \aj, 156, 264

\bibitem[{{Fulton} {et~al.}(2018){Fulton}, {Petigura}, {Blunt}, \&
  {Sinukoff}}]{Fulton18a}
{Fulton}, B.~J., {Petigura}, E.~A., {Blunt}, S., \& {Sinukoff}, E. 2018, \pasp,
  130, 044504

\bibitem[{{Gaia Collaboration} {et~al.}(2018){Gaia Collaboration}, {Brown},
  {Vallenari}, {Prusti}, {de Bruijne}, {Babusiaux}, {Bailer-Jones}, {Biermann},
  {Evans}, {Eyer}, {Jansen}, {Jordi}, {Klioner}, {Lammers}, {Lindegren},
  {Luri}, {Mignard}, {Panem}, {Pourbaix}, {Randich}, {Sartoretti}, {Siddiqui},
  {Soubiran}, {van Leeuwen}, {Walton}, {Arenou}, {Bastian}, {Cropper},
  {Drimmel}, {Katz}, {Lattanzi}, {Bakker}, {Cacciari}, {Casta{\~n}eda},
  {Chaoul}, {Cheek}, {De Angeli}, {Fabricius}, {Guerra}, {Holl}, {Masana},
  {Messineo}, {Mowlavi}, {Nienartowicz}, {Panuzzo}, {Portell}, {Riello},
  {Seabroke}, {Tanga}, {Th{\'e}venin}, {Gracia-Abril}, {Comoretto},
  {Garcia-Reinaldos}, {Teyssier}, {Altmann}, {Andrae}, {Audard},
  {Bellas-Velidis}, {Benson}, {Berthier}, {Blomme}, {Burgess}, {Busso},
  {Carry}, {Cellino}, {Clementini}, {Clotet}, {Creevey}, {Davidson}, {De
  Ridder}, {Delchambre}, {Dell'Oro}, {Ducourant},
  {Fern{\'a}ndez-Hern{\'a}ndez}, {Fouesneau}, {Fr{\'e}mat}, {Galluccio},
  {Garc{\'\i}a-Torres}, {Gonz{\'a}lez-N{\'u}{\~n}ez}, {Gonz{\'a}lez-Vidal},
  {Gosset}, {Guy}, {Halbwachs}, {Hambly}, {Harrison}, {Hern{\'a}ndez},
  {Hestroffer}, {Hodgkin}, {Hutton}, {Jasniewicz}, {Jean-Antoine-Piccolo},
  {Jordan}, {Korn}, {Krone-Martins}, {Lanzafame}, {Lebzelter}, {L{\"o}ffler},
  {Manteiga}, {Marrese}, {Mart{\'\i}n-Fleitas}, {Moitinho}, {Mora}, {Muinonen},
  {Osinde}, {Pancino}, {Pauwels}, {Petit}, {Recio-Blanco}, {Richards},
  {Rimoldini}, {Robin}, {Sarro}, {Siopis}, {Smith}, {Sozzetti}, {S{\"u}veges},
  {Torra}, {van Reeven}, {Abbas}, {Abreu Aramburu}, {Accart}, {Aerts},
  {Altavilla}, {{\'A}lvarez}, {Alvarez}, {Alves}, {Anderson}, {Andrei},
  {Anglada Varela}, {Antiche}, {Antoja}, {Arcay}, {Astraatmadja}, {Bach},
  {Baker}, {Balaguer-N{\'u}{\~n}ez}, {Balm}, {Barache}, {Barata}, {Barbato},
  {Barblan}, {Barklem}, {Barrado}, {Barros}, {Barstow}, {Bartholom{\'e}
  Mu{\~n}oz}, {Bassilana}, {Becciani}, {Bellazzini}, {Berihuete}, {Bertone},
  {Bianchi}, {Bienaym{\'e}}, {Blanco-Cuaresma}, {Boch}, {Boeche}, {Bombrun},
  {Borrachero}, {Bossini}, {Bouquillon}, {Bourda}, {Bragaglia}, {Bramante},
  {Breddels}, {Bressan}, {Brouillet}, {Br{\"u}semeister}, {Brugaletta},
  {Bucciarelli}, {Burlacu}, {Busonero}, {Butkevich}, {Buzzi}, {Caffau},
  {Cancelliere}, {Cannizzaro}, {Cantat-Gaudin}, {Carballo}, {Carlucci},
  {Carrasco}, {Casamiquela}, {Castellani}, {Castro-Ginard}, {Charlot},
  {Chemin}, {Chiavassa}, {Cocozza}, {Costigan}, {Cowell}, {Crifo}, {Crosta},
  {Crowley}, {Cuypers}, {Dafonte}, {Damerdji}, {Dapergolas}, {David}, {David},
  {de Laverny}, {De Luise}, {De March}, {de Martino}, {de Souza}, {de Torres},
  {Debosscher}, {del Pozo}, {Delbo}, {Delgado}, {Delgado}, {Di Matteo},
  {Diakite}, {Diener}, {Distefano}, {Dolding}, {Drazinos}, {Dur{\'a}n},
  {Edvardsson}, {Enke}, {Eriksson}, {Esquej}, {Eynard Bontemps}, {Fabre},
  {Fabrizio}, {Faigler}, {Falc{\~a}o}, {Farr{\`a}s Casas}, {Federici},
  {Fedorets}, {Fernique}, {Figueras}, {Filippi}, {Findeisen}, {Fonti},
  {Fraile}, {Fraser}, {Fr{\'e}zouls}, {Gai}, {Galleti}, {Garabato},
  {Garc{\'\i}a-Sedano}, {Garofalo}, {Garralda}, {Gavel}, {Gavras}, {Gerssen},
  {Geyer}, {Giacobbe}, {Gilmore}, {Girona}, {Giuffrida}, {Glass}, {Gomes},
  {Granvik}, {Gueguen}, {Guerrier}, {Guiraud}, {Guti{\'e}rrez-S{\'a}nchez},
  {Haigron}, {Hatzidimitriou}, {Hauser}, {Haywood}, {Heiter}, {Helmi}, {Heu},
  {Hilger}, {Hobbs}, {Hofmann}, {Holland}, {Huckle}, {Hypki}, {Icardi},
  {Jan{\ss}en}, {Jevardat de Fombelle}, {Jonker}, {Juh{\'a}sz}, {Julbe},
  {Karampelas}, {Kewley}, {Klar}, {Kochoska}, {Kohley}, {Kolenberg},
  {Kontizas}, {Kontizas}, {Koposov}, {Kordopatis}, {Kostrzewa-Rutkowska},
  {Koubsky}, {Lambert}, {Lanza}, {Lasne}, {Lavigne}, {Le Fustec}, {Le
  Poncin-Lafitte}, {Lebreton}, {Leccia}, {Leclerc}, {Lecoeur-Taibi},
  {Lenhardt}, {Leroux}, {Liao}, {Licata}, {Lindstr{\o}m}, {Lister}, {Livanou},
  {Lobel}, {L{\'o}pez}, {Managau}, {Mann}, {Mantelet}, {Marchal}, {Marchant},
  {Marconi}, {Marinoni}, {Marschalk{\'o}}, {Marshall}, {Martino}, {Marton},
  {Mary}, {Massari}, {Matijevi{\v{c}}}, {Mazeh}, {McMillan}, {Messina},
  {Michalik}, {Millar}, {Molina}, {Molinaro}, {Moln{\'a}r}, {Montegriffo},
  {Mor}, {Morbidelli}, {Morel}, {Morris}, {Mulone}, {Muraveva}, {Musella},
  {Nelemans}, {Nicastro}, {Noval}, {O'Mullane}, {Ord{\'e}novic},
  {Ord{\'o}{\~n}ez-Blanco}, {Osborne}, {Pagani}, {Pagano}, {Pailler},
  {Palacin}, {Palaversa}, {Panahi}, {Pawlak}, {Piersimoni}, {Pineau}, {Plachy},
  {Plum}, {Poggio}, {Poujoulet}, {Pr{\v{s}}a}, {Pulone}, {Racero}, {Ragaini},
  {Rambaux}, {Ramos-Lerate}, {Regibo}, {Reyl{\'e}}, {Riclet}, {Ripepi}, {Riva},
  {Rivard}, {Rixon}, {Roegiers}, {Roelens}, {Romero-G{\'o}mez}, {Rowell},
  {Royer}, {Ruiz-Dern}, {Sadowski}, {Sagrist{\`a} Sell{\'e}s}, {Sahlmann},
  {Salgado}, {Salguero}, {Sanna}, {Santana-Ros}, {Sarasso}, {Savietto},
  {Schultheis}, {Sciacca}, {Segol}, {Segovia}, {S{\'e}gransan}, {Shih},
  {Siltala}, {Silva}, {Smart}, {Smith}, {Solano}, {Solitro}, {Sordo}, {Soria
  Nieto}, {Souchay}, {Spagna}, {Spoto}, {Stampa}, {Steele},
  {Steidelm{\"u}ller}, {Stephenson}, {Stoev}, {Suess}, {Surdej}, {Szabados},
  {Szegedi-Elek}, {Tapiador}, {Taris}, {Tauran}, {Taylor}, {Teixeira},
  {Terrett}, {Teyssand ier}, {Thuillot}, {Titarenko}, {Torra Clotet}, {Turon},
  {Ulla}, {Utrilla}, {Uzzi}, {Vaillant}, {Valentini}, {Valette}, {van Elteren},
  {Van Hemelryck}, {van Leeuwen}, {Vaschetto}, {Vecchiato}, {Veljanoski},
  {Viala}, {Vicente}, {Vogt}, {von Essen}, {Voss}, {Votruba}, {Voutsinas},
  {Walmsley}, {Weiler}, {Wertz}, {Wevers}, {Wyrzykowski}, {Yoldas},
  {{\v{Z}}erjal}, {Ziaeepour}, {Zorec}, {Zschocke}, {Zucker}, {Zurbach}, \&
  {Zwitter}}]{GaiaDR2}
{Gaia Collaboration}, {Brown}, A.~G.~A., {Vallenari}, A., {et~al.} 2018, \aap,
  616, A1

\bibitem[{{Gelman} \& {Rubin}(1992)}]{Gelman92}
{Gelman}, A., \& {Rubin}, D.~B. 1992, Statistical Science, 7, 457

\bibitem[{{Goldreich} \& {Soter}(1966)}]{Goldreich66}
{Goldreich}, P., \& {Soter}, S. 1966, \icarus, 5, 375

\bibitem[{{Guerrero} {et~al.}(2021){Guerrero}, {Seager}, {Huang}, {Vanderburg},
  {Garcia Soto}, {Mireles}, {Hesse}, {Fong}, {Glidden}, {Shporer}, {Latham},
  {Collins}, {Quinn}, {Burt}, {Dragomir}, {Crossfield}, {Vanderspek},
  {Fausnaugh}, {Burke}, {Ricker}, {Daylan}, {Essack}, {G{\"u}nther}, {Osborn},
  {Pepper}, {Rowden}, {Sha}, {Villanueva}, {Yahalomi}, {Yu}, {Ballard},
  {Batalha}, {Berardo}, {Chontos}, {Dittmann}, {Esquerdo}, {Mikal-Evans},
  {Jayaraman}, {Krishnamurthy}, {Louie}, {Mehrle}, {Niraula}, {Rackham},
  {Rodriguez}, {Rowden}, {Sousa-Silva}, {Watanabe}, {Wong}, {Zhan},
  {Zivanovic}, {Christiansen}, {Ciardi}, {Swain}, {Lund}, {Mullally},
  {Fleming}, {Rodriguez}, {Boyd}, {Quintana}, {Barclay}, {Col{\'o}n},
  {Rinehart}, {Schlieder}, {Clampin}, {Jenkins}, {Twicken}, {Caldwell},
  {Coughlin}, {Henze}, {Lissauer}, {Morris}, {Rose}, {Smith}, {Tenenbaum},
  {Ting}, {Wohler}, {Bakos}, {Bean}, {Berta-Thompson}, {Bieryla}, {Bouma},
  {Buchhave}, {Butler}, {Charbonneau}, {Doty}, {Ge}, {Holman}, {Howard},
  {Kaltenegger}, {Kane}, {Kjeldsen}, {Kreidberg}, {Lin}, {Minsky}, {Narita},
  {Paegert}, {P{\'a}l}, {Palle}, {Sasselov}, {Spencer}, {Sozzetti}, {Stassun},
  {Torres}, {Udry}, \& {Winn}}]{Guerrero21}
{Guerrero}, N.~M., {Seager}, S., {Huang}, C.~X., {et~al.} 2021, \apjs, 254, 39

\bibitem[{{Hadden} \& {Lithwick}(2014)}]{Hadden14}
{Hadden}, S., \& {Lithwick}, Y. 2014, \apj, 787, 80

\bibitem[{{Hansen}(2017)}]{Hansen17}
{Hansen}, B. M.~S. 2017, \mnras, 467, 1531

\bibitem[{{Hara} {et~al.}(2017){Hara}, {Bou{\'e}}, {Laskar}, \&
  {Correia}}]{Hara17}
{Hara}, N.~C., {Bou{\'e}}, G., {Laskar}, J., \& {Correia}, A.~C.~M. 2017,
  \mnras, 464, 1220

\bibitem[{Harris {et~al.}(2020)Harris, Millman, van~der Walt, Gommers,
  Virtanen, Cournapeau, Wieser, Taylor, Berg, Smith, Kern, Picus, Hoyer, van
  Kerkwijk, Brett, Haldane, del R{\'{i}}o, Wiebe, Peterson,
  G{\'{e}}rard-Marchant, Sheppard, Reddy, Weckesser, Abbasi, Gohlke, \&
  Oliphant}]{harris2020array}
Harris, C.~R., Millman, K.~J., van~der Walt, S.~J., {et~al.} 2020, Nature, 585,
  357.
\newblock \url{https://doi.org/10.1038/s41586-020-2649-2}

\bibitem[{{Hinton}(2016)}]{Hinton16}
{Hinton}, S.~R. 2016, The Journal of Open Source Software, 1, 00045

\bibitem[{{Hodapp} {et~al.}(2003){Hodapp}, {Jensen}, {Irwin}, {Yamada},
  {Chung}, {Fletcher}, {Robertson}, {Hora}, {Simons}, {Mays}, {Nolan}, {Bec},
  {Merrill}, \& {Fowler}}]{Hodapp03}
{Hodapp}, K.~W., {Jensen}, J.~B., {Irwin}, E.~M., {et~al.} 2003, \pasp, 115,
  1388

\bibitem[{{Hoffman} \& {Gelman}(2011)}]{Hoffman11}
{Hoffman}, M.~D., \& {Gelman}, A. 2011, arXiv e-prints, arXiv:1111.4246

\bibitem[{{Howard} {et~al.}(2010){Howard}, {Johnson}, {Marcy}, {Fischer},
  {Wright}, {Bernat}, {Henry}, {Peek}, {Isaacson}, {Apps}, {Endl}, {Cochran},
  {Valenti}, {Anderson}, \& {Piskunov}}]{Howard10}
{Howard}, A.~W., {Johnson}, J.~A., {Marcy}, G.~W., {et~al.} 2010, \apj, 721,
  1467

\bibitem[{{Huber} {et~al.}(2017){Huber}, {Zinn}, {Bojsen-Hansen},
  {Pinsonneault}, {Sahlholdt}, {Serenelli}, {Silva Aguirre}, {Stassun},
  {Stello}, {Tayar}, {Bastien}, {Bedding}, {Buchhave}, {Chaplin}, {Davies},
  {Garc{\'\i}a}, {Latham}, {Mathur}, {Mosser}, \& {Sharma}}]{Huber17}
{Huber}, D., {Zinn}, J., {Bojsen-Hansen}, M., {et~al.} 2017, \apj, 844, 102

\bibitem[{Hunter(2007)}]{Hunter07}
Hunter, J.~D. 2007, Computing In Science \& Engineering, 9, 90

\bibitem[{{Isaacson} \& {Fischer}(2010)}]{Isaacson10}
{Isaacson}, H., \& {Fischer}, D. 2010, \apj, 725, 875

\bibitem[{{Izidoro} {et~al.}(2017){Izidoro}, {Ogihara}, {Raymond},
  {Morbidelli}, {Pierens}, {Bitsch}, {Cossou}, \& {Hersant}}]{Izidoro17}
{Izidoro}, A., {Ogihara}, M., {Raymond}, S.~N., {et~al.} 2017, \mnras, 470,
  1750

\bibitem[{{Jenkins}(2002)}]{Jenkins02}
{Jenkins}, J.~M. 2002, \apj, 575, 493

\bibitem[{{Jenkins} {et~al.}(2010){Jenkins}, {Chandrasekaran}, {McCauliff},
  {Caldwell}, {Tenenbaum}, {Li}, {Klaus}, {Cote}, \& {Middour}}]{Jenkins10}
{Jenkins}, J.~M., {Chandrasekaran}, H., {McCauliff}, S.~D., {et~al.} 2010, in
  Society of Photo-Optical Instrumentation Engineers (SPIE) Conference Series,
  Vol. 7740, Software and Cyberinfrastructure for Astronomy, ed. N.~M.
  {Radziwill} \& A.~{Bridger}, 77400D

\bibitem[{{Jenkins} {et~al.}(2016){Jenkins}, {Twicken}, {McCauliff},
  {Campbell}, {Sanderfer}, {Lung}, {Mansouri-Samani}, {Girouard}, {Tenenbaum},
  {Klaus}, {Smith}, {Caldwell}, {Chacon}, {Henze}, {Heiges}, {Latham},
  {Morgan}, {Swade}, {Rinehart}, \& {Vanderspek}}]{Jenkins16}
{Jenkins}, J.~M., {Twicken}, J.~D., {McCauliff}, S., {et~al.} 2016, in Society
  of Photo-Optical Instrumentation Engineers (SPIE) Conference Series, Vol.
  9913, Software and Cyberinfrastructure for Astronomy IV, 99133E

\bibitem[{{Juri{\'c}} \& {Tremaine}(2008)}]{Juric08}
{Juri{\'c}}, M., \& {Tremaine}, S. 2008, \apj, 686, 603

\bibitem[{{Kane} {et~al.}(2012){Kane}, {Ciardi}, {Gelino}, \& {von
  Braun}}]{Kane12}
{Kane}, S.~R., {Ciardi}, D.~R., {Gelino}, D.~M., \& {von Braun}, K. 2012,
  \mnras, 425, 757

\bibitem[{{Kipping}(2010)}]{Kipping10}
{Kipping}, D.~M. 2010, \mnras, 407, 301

\bibitem[{{Kipping}(2014)}]{Kipping14}
---. 2014, \mnras, 444, 2263

\bibitem[{{Kov{\'a}cs} {et~al.}(2002){Kov{\'a}cs}, {Zucker}, \&
  {Mazeh}}]{Kovacs02}
{Kov{\'a}cs}, G., {Zucker}, S., \& {Mazeh}, T. 2002, \aap, 391, 369

\bibitem[{{Lainey}(2016)}]{Lainey16}
{Lainey}, V. 2016, Celestial Mechanics and Dynamical Astronomy, 126, 145

\bibitem[{{Lee} \& {Thommes}(2009)}]{Lee09}
{Lee}, M.~H., \& {Thommes}, E.~W. 2009, \apj, 702, 1662

\bibitem[{{Li} {et~al.}(2019){Li}, {Tenenbaum}, {Twicken}, {Burke}, {Jenkins},
  {Quintana}, {Rowe}, \& {Seader}}]{Li19}
{Li}, J., {Tenenbaum}, P., {Twicken}, J.~D., {et~al.} 2019, \pasp, 131, 024506

\bibitem[{{Lightkurve Collaboration} {et~al.}(2018){Lightkurve Collaboration},
  {Cardoso}, {Hedges}, {Gully-Santiago}, {Saunders}, {Cody}, {Barclay}, {Hall},
  {Sagear}, {Turtelboom}, {Zhang}, {Tzanidakis}, {Mighell}, {Coughlin}, {Bell},
  {Berta-Thompson}, {Williams}, {Dotson}, \& {Barentsen}}]{lightkurve18}
{Lightkurve Collaboration}, {Cardoso}, J. V. d. M.~a., {Hedges}, C., {et~al.}
  2018, {Lightkurve: Kepler and TESS time series analysis in Python}, , ,
  ascl:1812.013

\bibitem[{{Lindegren} {et~al.}(2018){Lindegren}, {Hern{\'a}ndez}, {Bombrun},
  {Klioner}, {Bastian}, {Ramos-Lerate}, {de Torres}, {Steidelm{\"u}ller},
  {Stephenson}, {Hobbs}, {Lammers}, {Biermann}, {Geyer}, {Hilger}, {Michalik},
  {Stampa}, {McMillan}, {Casta{\~n}eda}, {Clotet}, {Comoretto}, {Davidson},
  {Fabricius}, {Gracia}, {Hambly}, {Hutton}, {Mora}, {Portell}, {van Leeuwen},
  {Abbas}, {Abreu}, {Altmann}, {Andrei}, {Anglada}, {Balaguer-N{\'u}{\~n}ez},
  {Barache}, {Becciani}, {Bertone}, {Bianchi}, {Bouquillon}, {Bourda},
  {Br{\"u}semeister}, {Bucciarelli}, {Busonero}, {Buzzi}, {Cancelliere},
  {Carlucci}, {Charlot}, {Cheek}, {Crosta}, {Crowley}, {de Bruijne}, {de
  Felice}, {Drimmel}, {Esquej}, {Fienga}, {Fraile}, {Gai}, {Garralda},
  {Gonz{\'a}lez-Vidal}, {Guerra}, {Hauser}, {Hofmann}, {Holl}, {Jordan},
  {Lattanzi}, {Lenhardt}, {Liao}, {Licata}, {Lister}, {L{\"o}ffler},
  {Marchant}, {Martin-Fleitas}, {Messineo}, {Mignard}, {Morbidelli}, {Poggio},
  {Riva}, {Rowell}, {Salguero}, {Sarasso}, {Sciacca}, {Siddiqui}, {Smart},
  {Spagna}, {Steele}, {Taris}, {Torra}, {van Elteren}, {van Reeven}, \&
  {Vecchiato}}]{Lindegren18}
{Lindegren}, L., {Hern{\'a}ndez}, J., {Bombrun}, A., {et~al.} 2018, \aap, 616,
  A2

\bibitem[{{Lopez} \& {Fortney}(2014)}]{Lopez14a}
{Lopez}, E.~D., \& {Fortney}, J.~J. 2014, \apj, 792, 1

\bibitem[{{Luhn} {et~al.}(2020){Luhn}, {Wright}, {Howard}, \&
  {Isaacson}}]{Luhn20}
{Luhn}, J.~K., {Wright}, J.~T., {Howard}, A.~W., \& {Isaacson}, H. 2020, \aj,
  159, 235

\bibitem[{{Mandel} \& {Agol}(2002)}]{Mandel02}
{Mandel}, K., \& {Agol}, E. 2002, \apjl, 580, L171

\bibitem[{{Marcy} \& {Butler}(1992)}]{Marcy92}
{Marcy}, G.~W., \& {Butler}, R.~P. 1992, \pasp, 104, 270

\bibitem[{{Marcy} {et~al.}(2014){Marcy}, {Isaacson}, {Howard}, {Rowe},
  {Jenkins}, {Bryson}, {Latham}, {Howell}, {Gautier}, {Batalha}, {Rogers},
  {Ciardi}, {Fischer}, {Gilliland}, {Kjeldsen}, {Christensen-Dalsgaard},
  {Huber}, {Chaplin}, {Basu}, {Buchhave}, {Quinn}, {Borucki}, {Koch}, {Hunter},
  {Caldwell}, {Van Cleve}, {Kolbl}, {Weiss}, {Petigura}, {Seager}, {Morton},
  {Johnson}, {Ballard}, {Burke}, {Cochran}, {Endl}, {MacQueen}, {Everett},
  {Lissauer}, {Ford}, {Torres}, {Fressin}, {Brown}, {Steffen}, {Charbonneau},
  {Basri}, {Sasselov}, {Winn}, {Sanchis-Ojeda}, {Christiansen}, {Adams},
  {Henze}, {Dupree}, {Fabrycky}, {Fortney}, {Tarter}, {Holman}, {Tenenbaum},
  {Shporer}, {Lucas}, {Welsh}, {Orosz}, {Bedding}, {Campante}, {Davies},
  {Elsworth}, {Handberg}, {Hekker}, {Karoff}, {Kawaler}, {Lund}, {Lundkvist},
  {Metcalfe}, {Miglio}, {Silva Aguirre}, {Stello}, {White}, {Boss}, {Devore},
  {Gould}, {Prsa}, {Agol}, {Barclay}, {Coughlin}, {Brugamyer}, {Mullally},
  {Quintana}, {Still}, {Thompson}, {Morrison}, {Twicken}, {D{\'e}sert},
  {Carter}, {Crepp}, {H{\'e}brard}, {Santerne}, {Moutou}, {Sobeck}, {Hudgins},
  {Haas}, {Robertson}, {Lillo-Box}, \& {Barrado}}]{Marcy14}
{Marcy}, G.~W., {Isaacson}, H., {Howard}, A.~W., {et~al.} 2014, \apjs, 210, 20

\bibitem[{{Mardling} \& {Lin}(2004)}]{Mardling04}
{Mardling}, R.~A., \& {Lin}, D.~N.~C. 2004, \apj, 614, 955

\bibitem[{McKinney(2010)}]{pandas}
McKinney, W. 2010, in Proceedings of the 9th Python in Science Conference, ed.
  S.~van~der Walt \& J.~Millman, 51 -- 56

\bibitem[{{Moorhead} {et~al.}(2011){Moorhead}, {Ford}, {Morehead}, {Rowe},
  {Borucki}, {Batalha}, {Bryson}, {Caldwell}, {Fabrycky}, {Gautier}, {Koch},
  {Holman}, {Jenkins}, {Li}, {Lissauer}, {Lucas}, {Marcy}, {Quinn}, {Quintana},
  {Ragozzine}, {Shporer}, {Still}, \& {Torres}}]{Moorhead11}
{Moorhead}, A.~V., {Ford}, E.~B., {Morehead}, R.~C., {et~al.} 2011, \apjs, 197,
  1

\bibitem[{{Murray} \& {Dermott}(1999)}]{Murray99}
{Murray}, C.~D., \& {Dermott}, S.~F. 1999, {Solar system dynamics}

\bibitem[{{Naoz}(2016)}]{Naoz16}
{Naoz}, S. 2016, \araa, 54, 441

\bibitem[{{Osborn} {et~al.}(2017){Osborn}, {Santerne}, {Barros}, {Santos},
  {Dumusque}, {Malavolta}, {Armstrong}, {Hojjatpanah}, {Demangeon},
  {Adibekyan}, {Almenara}, {Barrado}, {Bayliss}, {Boisse}, {Bouchy}, {Brown},
  {Cameron}, {Charbonneau}, {Deleuil}, {Delgado Mena}, {D{\'\i}az},
  {H{\'e}brard}, {Kirk}, {King}, {Lam}, {Latham}, {Lillo-Box}, {Louden},
  {Lovis}, {Marmier}, {McCormac}, {Molinari}, {Pepe}, {Pollacco}, {Sousa},
  {Udry}, \& {Walker}}]{Osborn17}
{Osborn}, H.~P., {Santerne}, A., {Barros}, S.~C.~C., {et~al.} 2017, \aap, 604,
  A19

\bibitem[{{Owen} \& {Wu}(2017)}]{Owen17}
{Owen}, J.~E., \& {Wu}, Y. 2017, \apj, 847, 29

\bibitem[{{Parviainen} \& {Aigrain}(2015)}]{ldtk15}
{Parviainen}, H., \& {Aigrain}, S. 2015, \mnras, 453, 3821

\bibitem[{{Petigura}(2020)}]{Petigura20}
{Petigura}, E.~A. 2020, \aj, 160, 89

\bibitem[{{Petigura} {et~al.}(2017{\natexlab{a}}){Petigura}, {Howard}, {Marcy},
  {Johnson}, {Isaacson}, {Cargile}, {Hebb}, {Fulton}, {Weiss}, {Morton},
  {Winn}, {Rogers}, {Sinukoff}, {Hirsch}, \& {Crossfield}}]{Petigura17b}
{Petigura}, E.~A., {Howard}, A.~W., {Marcy}, G.~W., {et~al.}
  2017{\natexlab{a}}, \aj, 154, 107

\bibitem[{{Petigura} {et~al.}(2017{\natexlab{b}}){Petigura}, {Sinukoff},
  {Lopez}, {Crossfield}, {Howard}, {Brewer}, {Fulton}, {Isaacson}, {Ciardi},
  {Howell}, {Everett}, {Horch}, {Hirsch}, {Weiss}, \& {Schlieder}}]{Petigura17}
{Petigura}, E.~A., {Sinukoff}, E., {Lopez}, E.~D., {et~al.} 2017{\natexlab{b}},
  \aj, 153, 142

\bibitem[{{Press} {et~al.}(1992){Press}, {Teukolsky}, {Vetterling}, \&
  {Flannery}}]{Press92}
{Press}, W.~H., {Teukolsky}, S.~A., {Vetterling}, W.~T., \& {Flannery}, B.~P.
  1992, {Numerical recipes in C. The art of scientific computing}

\bibitem[{{Pu} \& {Lai}(2018)}]{Pu18}
{Pu}, B., \& {Lai}, D. 2018, \mnras, 478, 197

\bibitem[{{Rein} \& {Liu}(2012)}]{Rein12}
{Rein}, H., \& {Liu}, S.~F. 2012, \aap, 537, A128

\bibitem[{{Ricker} {et~al.}(2015){Ricker}, {Winn}, {Vanderspek}, {Latham},
  {Bakos}, {Bean}, {Berta-Thompson}, {Brown}, {Buchhave}, {Butler}, {Butler},
  {Chaplin}, {Charbonneau}, {Christensen-Dalsgaard}, {Clampin}, {Deming},
  {Doty}, {De Lee}, {Dressing}, {Dunham}, {Endl}, {Fressin}, {Ge}, {Henning},
  {Holman}, {Howard}, {Ida}, {Jenkins}, {Jernigan}, {Johnson}, {Kaltenegger},
  {Kawai}, {Kjeldsen}, {Laughlin}, {Levine}, {Lin}, {Lissauer}, {MacQueen},
  {Marcy}, {McCullough}, {Morton}, {Narita}, {Paegert}, {Palle}, {Pepe},
  {Pepper}, {Quirrenbach}, {Rinehart}, {Sasselov}, {Sato}, {Seager},
  {Sozzetti}, {Stassun}, {Sullivan}, {Szentgyorgyi}, {Torres}, {Udry}, \&
  {Villasenor}}]{Ricker15}
{Ricker}, G.~R., {Winn}, J.~N., {Vanderspek}, R., {et~al.} 2015, Journal of
  Astronomical Telescopes, Instruments, and Systems, 1, 014003

\bibitem[{{Rogers} \& {Showman}(2014)}]{Rogers14}
{Rogers}, T.~M., \& {Showman}, A.~P. 2014, \apjl, 782, L4

\bibitem[{{Rosenthal} {et~al.}(2021){Rosenthal}, {Fulton}, {Hirsch},
  {Isaacson}, {Howard}, {Dedrick}, {Sherstyuk}, {Blunt}, {Petigura}, {Knutson},
  {Behmard}, {Chontos}, {Crepp}, {Crossfield}, {Dalba}, {Fischer}, {Henry},
  {Kane}, {Kosiarek}, {Marcy}, {Rubenzahl}, {Weiss}, \& {Wright}}]{Rosenthal21}
{Rosenthal}, L.~J., {Fulton}, B.~J., {Hirsch}, L.~A., {et~al.} 2021, \apjs,
  255, 8

\bibitem[{{Salvatier} {et~al.}(2016){Salvatier}, {Wiecki}, \&
  {Fonnesbeck}}]{pymc16}
{Salvatier}, J., {Wiecki}, T.~V., \& {Fonnesbeck}, C. 2016, {PyMC3: Python
  probabilistic programming framework}, , , ascl:1610.016

\bibitem[{{Sandford} \& {Kipping}(2017)}]{Sandford17}
{Sandford}, E., \& {Kipping}, D. 2017, \aj, 154, 228

\bibitem[{{Schwarz}(1978)}]{Schwarz78}
{Schwarz}, G. 1978, Annals of Statistics, 6, 461

\bibitem[{{Skrutskie} {et~al.}(2006){Skrutskie}, {Cutri}, {Stiening},
  {Weinberg}, {Schneider}, {Carpenter}, {Beichman}, {Capps}, {Chester},
  {Elias}, {Huchra}, {Liebert}, {Lonsdale}, {Monet}, {Price}, {Seitzer},
  {Jarrett}, {Kirkpatrick}, {Gizis}, {Howard}, {Evans}, {Fowler}, {Fullmer},
  {Hurt}, {Light}, {Kopan}, {Marsh}, {McCallon}, {Tam}, {Van Dyk}, \&
  {Wheelock}}]{Skrutskie06}
{Skrutskie}, M.~F., {Cutri}, R.~M., {Stiening}, R., {et~al.} 2006, \aj, 131,
  1163

\bibitem[{{Smith} {et~al.}(2012){Smith}, {Stumpe}, {Van Cleve}, {Jenkins},
  {Barclay}, {Fanelli}, {Girouard}, {Kolodziejczak}, {McCauliff}, {Morris}, \&
  {Twicken}}]{Smith12}
{Smith}, J.~C., {Stumpe}, M.~C., {Van Cleve}, J.~E., {et~al.} 2012, \pasp, 124,
  1000

\bibitem[{{Stassun} {et~al.}(2019){Stassun}, {Oelkers}, {Paegert}, {Torres},
  {Pepper}, {De Lee}, {Collins}, {Latham}, {Muirhead}, {Chittidi},
  {Rojas-Ayala}, {Fleming}, {Rose}, {Tenenbaum}, {Ting}, {Kane}, {Barclay},
  {Bean}, {Brassuer}, {Charbonneau}, {Ge}, {Lissauer}, {Mann}, {McLean},
  {Mullally}, {Narita}, {Plavchan}, {Ricker}, {Sasselov}, {Seager}, {Sharma},
  {Shiao}, {Sozzetti}, {Stello}, {Vanderspek}, {Wallace}, \&
  {Winn}}]{Stassun19}
{Stassun}, K.~G., {Oelkers}, R.~J., {Paegert}, M., {et~al.} 2019, \aj, 158, 138

\bibitem[{{Stumpe} {et~al.}(2014){Stumpe}, {Smith}, {Catanzarite}, {Van Cleve},
  {Jenkins}, {Twicken}, \& {Girouard}}]{Stumpe14}
{Stumpe}, M.~C., {Smith}, J.~C., {Catanzarite}, J.~H., {et~al.} 2014, \pasp,
  126, 100

\bibitem[{{Stumpe} {et~al.}(2012){Stumpe}, {Smith}, {Van Cleve}, {Twicken},
  {Barclay}, {Fanelli}, {Girouard}, {Jenkins}, {Kolodziejczak}, {McCauliff}, \&
  {Morris}}]{Stumpe12}
{Stumpe}, M.~C., {Smith}, J.~C., {Van Cleve}, J.~E., {et~al.} 2012, \pasp, 124,
  985

\bibitem[{{Tayar} {et~al.}(2020){Tayar}, {Claytor}, {Huber}, \& {van
  Saders}}]{Tayar20}
{Tayar}, J., {Claytor}, Z.~R., {Huber}, D., \& {van Saders}, J. 2020, arXiv
  e-prints, arXiv:2012.07957

\bibitem[{{The Theano Development Team} {et~al.}(2016){The Theano Development
  Team}, {Al-Rfou}, {Alain}, {Almahairi}, {Angermueller}, {Bahdanau}, {Ballas},
  {Bastien}, {Bayer}, {Belikov}, {Belopolsky}, {Bengio}, {Bergeron},
  {Bergstra}, {Bisson}, {Bleecher Snyder}, {Bouchard}, {Boulanger-Lewandowski},
  {Bouthillier}, {de Br{\'e}bisson}, {Breuleux}, {Carrier}, {Cho}, {Chorowski},
  {Christiano}, {Cooijmans}, {C{\^o}t{\'e}}, {C{\^o}t{\'e}}, {Courville},
  {Dauphin}, {Delalleau}, {Demouth}, {Desjardins}, {Dieleman}, {Dinh},
  {Ducoffe}, {Dumoulin}, {Ebrahimi Kahou}, {Erhan}, {Fan}, {Firat}, {Germain},
  {Glorot}, {Goodfellow}, {Graham}, {Gulcehre}, {Hamel}, {Harlouchet}, {Heng},
  {Hidasi}, {Honari}, {Jain}, {Jean}, {Jia}, {Korobov}, {Kulkarni}, {Lamb},
  {Lamblin}, {Larsen}, {Laurent}, {Lee}, {Lefrancois}, {Lemieux},
  {L{\'e}onard}, {Lin}, {Livezey}, {Lorenz}, {Lowin}, {Ma}, {Manzagol},
  {Mastropietro}, {McGibbon}, {Memisevic}, {van Merri{\"e}nboer}, {Michalski},
  {Mirza}, {Orlandi}, {Pal}, {Pascanu}, {Pezeshki}, {Raffel}, {Renshaw},
  {Rocklin}, {Romero}, {Roth}, {Sadowski}, {Salvatier}, {Savard},
  {Schl{\"u}ter}, {Schulman}, {Schwartz}, {Vlad Serban}, {Serdyuk},
  {Shabanian}, {Simon}, {Spieckermann}, {Ramana Subramanyam}, {Sygnowski},
  {Tanguay}, {van Tulder}, {Turian}, {Urban}, {Vincent}, {Visin}, {de Vries},
  {Warde-Farley}, {Webb}, {Willson}, {Xu}, {Xue}, {Yao}, {Zhang}, \&
  {Zhang}}]{theano16}
{The Theano Development Team}, {Al-Rfou}, R., {Alain}, G., {et~al.} 2016, arXiv
  e-prints, arXiv:1605.02688

\bibitem[{{Tsiganis} {et~al.}(2005){Tsiganis}, {Gomes}, {Morbidelli}, \&
  {Levison}}]{Tsiganis05}
{Tsiganis}, K., {Gomes}, R., {Morbidelli}, A., \& {Levison}, H.~F. 2005, \nat,
  435, 459

\bibitem[{{Twicken} {et~al.}(2018){Twicken}, {Catanzarite}, {Clarke},
  {Girouard}, {Jenkins}, {Klaus}, {Li}, {McCauliff}, {Seader}, {Tenenbaum},
  {Wohler}, {Bryson}, {Burke}, {Caldwell}, {Haas}, {Henze}, \&
  {Sanderfer}}]{Twicken18}
{Twicken}, J.~D., {Catanzarite}, J.~H., {Clarke}, B.~D., {et~al.} 2018, \pasp,
  130, 064502

\bibitem[{{Van Eylen} \& {Albrecht}(2015)}]{VE15}
{Van Eylen}, V., \& {Albrecht}, S. 2015, \apj, 808, 126

\bibitem[{{Van Eylen} {et~al.}(2014){Van Eylen}, {Lund}, {Silva Aguirre},
  {Arentoft}, {Kjeldsen}, {Albrecht}, {Chaplin}, {Isaacson}, {Pedersen},
  {Jessen-Hansen}, {Tingley}, {Christensen-Dalsgaard}, {Aerts}, {Campante}, \&
  {Bryson}}]{VE14}
{Van Eylen}, V., {Lund}, M.~N., {Silva Aguirre}, V., {et~al.} 2014, \apj, 782,
  14

\bibitem[{{Van Eylen} {et~al.}(2019){Van Eylen}, {Albrecht}, {Huang},
  {MacDonald}, {Dawson}, {Cai}, {Foreman-Mackey}, {Lundkvist}, {Silva Aguirre},
  {Snellen}, \& {Winn}}]{VE19}
{Van Eylen}, V., {Albrecht}, S., {Huang}, X., {et~al.} 2019, \aj, 157, 61

\bibitem[{{Virtanen} {et~al.}(2020){Virtanen}, {Gommers}, {Oliphant},
  {Haberland}, {Reddy}, {Cournapeau}, {Burovski}, {Peterson}, {Weckesser},
  {Bright}, {van der Walt}, {Brett}, {Wilson}, {Millman}, {Mayorov}, {Nelson},
  {Jones}, {Kern}, {Larson}, {Carey}, {Polat}, {Feng}, {Moore}, {Vand erPlas},
  {Laxalde}, {Perktold}, {Cimrman}, {Henriksen}, {Quintero}, {Harris},
  {Archibald}, {Ribeiro}, {Pedregosa}, {van Mulbregt}, \& {SciPy 1. 0
  Contributors}}]{SciPy20}
{Virtanen}, P., {Gommers}, R., {Oliphant}, T.~E., {et~al.} 2020, Nature
  Methods, 17, 261

\bibitem[{{Vogt} {et~al.}(1994){Vogt}, {Allen}, {Bigelow}, {Bresee}, {Brown},
  {Cantrall}, {Conrad}, {Couture}, {Delaney}, {Epps}, {Hilyard}, {Hilyard},
  {Horn}, {Jern}, {Kanto}, {Keane}, {Kibrick}, {Lewis}, {Osborne},
  {Pardeilhan}, {Pfister}, {Ricketts}, {Robinson}, {Stover}, {Tucker}, {Ward},
  \& {Wei}}]{Vogt94}
{Vogt}, S.~S., {Allen}, S.~L., {Bigelow}, B.~C., {et~al.} 1994, in Society of
  Photo-Optical Instrumentation Engineers (SPIE) Conference Series, Vol. 2198,
  Instrumentation in Astronomy VIII, ed. D.~L. {Crawford} \& E.~R. {Craine},
  362

\bibitem[{{Vogt} {et~al.}(2014){Vogt}, {Radovan}, {Kibrick}, {Butler},
  {Alcott}, {Allen}, {Arriagada}, {Bolte}, {Burt}, {Cabak}, {Chloros},
  {Cowley}, {Deich}, {Dupraw}, {Earthman}, {Epps}, {Faber}, {Fischer}, {Gates},
  {Hilyard}, {Holden}, {Johnston}, {Keiser}, {Kanto}, {Katsuki}, {Laiterman},
  {Lanclos}, {Laughlin}, {Lewis}, {Lockwood}, {Lynam}, {Marcy}, {McLean},
  {Miller}, {Misch}, {Peck}, {Pfister}, {Phillips}, {Rivera}, {Sand ford},
  {Saylor}, {Stover}, {Thompson}, {Walp}, {Ward}, {Wareham}, {Wei}, \&
  {Wright}}]{Vogt14}
{Vogt}, S.~S., {Radovan}, M., {Kibrick}, R., {et~al.} 2014, \pasp, 126, 359

\bibitem[{{Weiss} \& {Marcy}(2014)}]{Weiss14}
{Weiss}, L.~M., \& {Marcy}, G.~W. 2014, \apjl, 783, L6

\bibitem[{{Weiss} {et~al.}(2018){Weiss}, {Marcy}, {Petigura}, {Fulton},
  {Howard}, {Winn}, {Isaacson}, {Morton}, {Hirsch}, {Sinukoff}, {Cumming},
  {Hebb}, \& {Cargile}}]{Weiss18}
{Weiss}, L.~M., {Marcy}, G.~W., {Petigura}, E.~A., {et~al.} 2018, \aj, 155, 48

\bibitem[{{Winn}(2010)}]{Winn10}
{Winn}, J.~N. 2010, {Exoplanet Transits and Occultations}, ed. S.~{Seager},
  55--77

\bibitem[{{Winn} \& {Fabrycky}(2015)}]{Winn15}
{Winn}, J.~N., \& {Fabrycky}, D.~C. 2015, \araa, 53, 409

\bibitem[{{Wolfgang} {et~al.}(2016){Wolfgang}, {Rogers}, \&
  {Ford}}]{Wolfgang16}
{Wolfgang}, A., {Rogers}, L.~A., \& {Ford}, E.~B. 2016, \apj, 825, 19

\bibitem[{{Xie} {et~al.}(2016){Xie}, {Dong}, {Zhu}, {Huber}, {Zheng}, {De Cat},
  {Fu}, {Liu}, {Luo}, {Wu}, {Zhang}, {Zhang}, {Zhou}, {Cao}, {Hou}, {Wang}, \&
  {Zhang}}]{Xie16}
{Xie}, J.-W., {Dong}, S., {Zhu}, Z., {et~al.} 2016, Proceedings of the National
  Academy of Science, 113, 11431

\bibitem[{{Zeng} {et~al.}(2016){Zeng}, {Sasselov}, \& {Jacobsen}}]{Zeng16}
{Zeng}, L., {Sasselov}, D.~D., \& {Jacobsen}, S.~B. 2016, \apj, 819, 127

\bibitem[{{Zeng} {et~al.}(2019){Zeng}, {Jacobsen}, {Sasselov}, {Petaev},
  {Vanderburg}, {Lopez-Morales}, {Perez-Mercader}, {Mattsson}, {Li}, {Heising},
  {Bonomo}, {Damasso}, {Berger}, {Cao}, {Levi}, \& {Wordsworth}}]{Zeng19}
{Zeng}, L., {Jacobsen}, S.~B., {Sasselov}, D.~D., {et~al.} 2019, Proceedings of
  the National Academy of Science, 116, 9723

\end{thebibliography}

\clearpage

\end{document}